%% file: Main.tex
\documentclass[journal]{IEEEtran}
\usepackage{forest}
\usepackage{float}
\usepackage{cellspace}
\usepackage{diagbox}
 \usepackage{comment}
\usepackage{longtable}

\usepackage{multirow}
\usepackage{tabularx}
\usepackage{makecell}
\usepackage{xcolor}

\usepackage[nopostdot,style=super,nonumberlist,toc,nogroupskip]{glossaries}
\usepackage{tikz} 
\usetikzlibrary{shapes.geometric, arrows,trees}
\usepackage{tikz-qtree}
\usetikzlibrary{trees}
 \usetikzlibrary{shapes.geometric, arrows}
\usepackage{cite}
\usepackage{amssymb}
\usepackage[english]{babel}
\usepackage{graphicx}
\usepackage{colortbl}
\usepackage{csquotes}
\usepackage{array,etoolbox}
\usepackage[utf8]{inputenc}
\usepackage{fontawesome}
\usepackage{caption}
\usepackage{subcaption}
\usepackage{amsmath}
\usepackage{glossaries}

\newcommand{\eg}{\textit{e}.\textit{g}.}

\usepackage{enumitem}
\newlist{tabitemize}{itemize}{1}
\setlist[tabitemize]{nosep,
                  topsep= 0pt,
                  partopsep=0pt,
                  leftmargin= *,
                  label=\textbullet,
                  before=\vspace{0.3\baselineskip},
                  after=\vspace{-\baselineskip}
                  }

\newcolumntype{M}[1]{>{\centering\arraybackslash}m{#1}}
\makeglossaries
 \usepackage[printonlyused]{acronym}
\usepackage{booktabs}

\makeatletter
\renewenvironment{AC@deflist}[1]%
{\ifAC@nolist\else%
    \raggedright\begin{list}{}%
        {%
            \setlength{\labelwidth}{#1}%
            \setlength{\leftmargin}{\labelwidth}%
            \addtolength{\leftmargin}{\labelsep}%
            \renewcommand{\makelabel}{\AC@makelabel}}%
        \fi}%
    {\ifAC@nolist\else\end{list}\fi}%
\def\Acro#1[#2]#3{%
    \vspace{-\normalbaselineskip}\acronym[\dimexpr0.2\textwidth]\acro{#1}[#2]{#3}\endacronym
     \vspace{-2\normalbaselineskip}}
\makeatother
\usepackage{tikz}
\usepackage{array,etoolbox}
\usepackage[utf8]{inputenc}
\makeglossaries
\setacronymstyle{long-short}
\input{acro2}

\begin{document}
 
\title{A Comprehensive Survey on 6G Networks: Applications, Core Services, Enabling Technologies, and Future Challenges}
 
\author{Amin~Shahraki\textsuperscript{*},~\IEEEmembership{Member,~IEEE,}
        Mahmoud~Abbasi,~\IEEEmembership{Member,~IEEE,} ~Md.~Jalil Piran\textsuperscript{*},~\IEEEmembership{Senior Member,~IEEE}~and~ Amir Taherkordi% <-this % stops a space
       \thanks{Amin Shahraki (\textit{Corresponding Author}) is with the  Department of Informatics, University of Oslo, Oslo, Norway (e-mail: am.shahraki@ieee.org)}
       \thanks{Mahmoud Abbasi is with Islamic Azad University, Mashhad, Iran (email: mahmoud.abbasi@ieee.og)}
% <-this % stops a space
\thanks{Md. Jalil Piran (\textit{Corresponding Author}) is with the Department of Computer Science and Engineering, Sejong University, Seoul, 05006, South Korea (email: piran@sejong.ac.kr)}
\thanks{Amir Taherkordi is with University of Oslo, Oslo, Norway (email: amirhost@ifi.uio.no)}% <-this % stops a space
\thanks{Manuscript received xxx xx, 2021; revised xxx xx, 2021.}
}

\markboth{IEEE Transactions on Network and Service Management,~Vol.~xx, No.~xx, xx~2021}%
{Abbasi \MakeLowercase{\textit{et al.}}: 6G Wireless Networks}
 
\maketitle
\begingroup\renewcommand\thefootnote{*}
\footnotetext{Equal contribution in case of corresponding author}
\endgroup
\textcolor{blue}{[\textit{Note: This work has been submitted to the IEEE Transactions on Network and Service Management journal for possible publication. Copyright may be transferred without notice, after which this version may no longer be accessible }]}\\
\begin{abstract}
\textcolor{black}{Cellular Internet of Things (IoT) is considered as \textit{de facto} paradigm to improve the communication and computation systems. Cellular IoT connects massive number of physical and virtual objects to the Internet using cellular networks.  
The latest generation of cellular networks, e.g. fifth-generation (5G), use evolutionary and revolutionary technologies to notably improve the performance of wireless networks. However, given the envisioned new use-cases, e.g., holographic communication, and the ever-increasing deployment of massive smart-physical end-devices in IoT, the volume of network traffic has considerably raised, and therefore, the current generation of mobile networks cannot wholly meet the ever-increasing demands. Hence, it is envisioned that the next generation, sixth generation (6G) networks, need to play a critical role to alleviate such challenges in IoT by providing new communication services, network capacity, and ultra-low latency communications (uRLLC). In this paper, first, the need for 6G networks is discussed. Then, the potential 6G requirements and trends, as well as the latest research activities related to 6G are introduced e.g., Tactile Internet and Terahertz (THz). Furthermore, the key performance indicators, applications, new services, and the potential key enabling technologies for 6G networks are presented. Finally, several potential unresolved challenges for future 6G networks are presented.}
\end{abstract}

 
\IEEEpeerreviewmaketitle

\input{Sections/Introduction}

\input{Sections/related_Work}

\input{Sections/requirements}

\input{Sections/Research_Activities_Motivation}

\input{Sections/KPIs_Use_Cases}

\input{Sections/New_Service_Classes}

\input{Sections/Evolutionary_technologies}

\input{Sections/Revolutionary_technologies}

\input{Sections/Future_Challenges}

\input{Sections/Conclusion}

\ifCLASSOPTIONcaptionsoff
  \newpage
\fi
 
\bibliographystyle{IEEEtran}
\bibliography{Main}
 
%\end{IEEEbiography}
 
\end{document}

%% file: acro2.tex
\newacronym{3cls}{3CLS}{Control, Localization, and Sensing}
\newacronym{3gpp}{3GPP}{3rd generation partnership project}
\newacronym{3d}{3D}{3-Dimensional}
\newacronym{ai}{AI}{artificial intelligence}
\newacronym{ar}{AR}{augmented reality}
\newacronym{aid}{AID}{Autonomous intelligent driving}
\newacronym{au}{AU}{aerial user}
\newacronym{aes}{AES}{Advanced Encryption Standard}
\newacronym{amp}{AMP}{approximate message passing}
\newacronym{ap}{APs}{access points}
\newacronym{cad}{CAD}{cooperative automated driving}
\newacronym{csi}{CSI}{channel state information}
\newacronym{cran}{CRAN}{cloud radio access network}
\newacronym{dl}{DL}{deep learning}
\newacronym{d2d}{D2D}{Device-to-device}
\newacronym{embb}{eMBB}{enhanced mobile broadband}
\newacronym{ecc}{ECC}{Elliptic Curve Encryption}
\newacronym{em}{EM}{emit electromagnetic}
\newacronym{ei}{EI}{Edge Intelligence}
\newacronym{e2e}{E2E}{end-to-end}
\newacronym{etsi}{ETSI}{European Telecommunications Standards Institute}
\newacronym{gps}{GPS}{global positioning system}
\newacronym{hmimos}{HMIMOS}{holographic MIMO surface}
\newacronym{iot}{IoT}{Internet of things}
\newacronym{iiot}{IIoT}{industrial \gls{iot}}
\newacronym{itu}{ITU}{International Telecommunication Union}
\newacronym{ict}{ICT}{Information and Communication Technology}
\newacronym{ids}{IDS}{intrusion detection system}
\newacronym{ir}{IR}{infrared}
\newacronym{kpi}{KPI}{key performance indicator}
\newacronym{los}{LoS}{line-of-sight}
\newacronym{lorawan}{LoRaWAN}{Long Range Wide Area Network}
\newacronym{lis}{LIS}{Large Intelligent Surfaces}
\newacronym{mimo}{MIMO}{multiple-input and multiple-output }
\newacronym{mtc}{MTC}{machine-type communication}
\newacronym{mmtc}{mMTC}{massive \gls{mtc}}
\newacronym{mmi}{MMI}{mind-machine interface}
\newacronym{mmwave}{mmWave}{millimetre wave}
\newacronym{ml}{ML}{machine learning}
\newacronym{mec}{MEC}{mobile edge computing}
\newacronym{nfv}{NFV}{network function virtualization}
\newacronym{nmo}{NMO}{Network Management and Orchestration}
\newacronym{ntc}{NTC}{network traffic classification}
\newacronym{nbiot}{NB-IoT}{Narrow band \gls{iot}}
\newacronym{ntp}{NTP}{network traffic prediction}
\newacronym{ntma}{NTMA}{network traffic monitoring and analysis}
\newacronym{noma}{NOMA}{Non-orthogonal multiple access}
\newacronym{ofdm}{OFDM}{Orthogonal frequency-division multiplexing}
\newacronym{owc}{OWC}{optical wireless communication}
\newacronym{oma}{OMA}{orthogonal multiple access}
\newacronym{plmn}{PLMN}{public land mobile network}
\newacronym{qos}{QoS}{quality of service}
\newacronym{qoe}{QoE}{quality of experience}
\newacronym{qc}{QC}{quantum computing}
\newacronym{qoc}{QOC}{quantum optical communication}
\newacronym{qkd}{QKD}{quantum key distribution}
\newacronym{ris}{RIS}{reconfigurable intelligent surfaces}
\newacronym{rl}{RL}{reinforcement learning}
\newacronym{ra}{RA}{Random Access}
\newacronym{rf}{RF}{radio frequency}
\newacronym{sdn}{SDN}{software defined networking}
\newacronym{sr}{SR}{super resolution}
\newacronym{swap}{SWAP}{size/weight and power}
\newacronym{sal}{SAL}{single-anchor localization}
\newacronym{svm}{SVM}{support vector machine}
\newacronym{sic}{SIC}{successive interference cancellation}
\newacronym{thz}{THz}{Terahertz}
\newacronym{urllc}{URLLC}{ultra-reliable low-latency communications}
\newacronym{uhsllc}{uHSLLC}{ultra-high speed-with-low-latency communication }
\newacronym{umub}{uMUB}{ubiquitous mobile ultra-broadband}
\newacronym{uhdd}{UHDD}{ultra-high data density}
\newacronym{uav}{UAV}{unmanned aerial vehicle}
\newacronym{ue}{UE}{User Equipment}
\newacronym{uv}{UV}{ultraviolet}
\newacronym{udhn}{UDHN}{ultra-dense heterogeneous networks}
\newacronym{v2x}{V2X}{vehicle to everything}
\newacronym{vr}{VR}{virtual reality}
\newacronym{vmr}{VMR}{virtual meeting room}
\newacronym{vlc}{VLC}{visible light communication}
\newacronym{wipt}{WIPT}{wireless information and power transfer}
\newacronym{wpt}{WPT}{wireless power transfer}
\newacronym{xr}{XR}{extended reality}

%% file: Sections/Introduction.tex
\section{Introduction}\label{Introduction}

\textcolor{black}{\gls{iot} is the most prevalent framework in the realm of \gls{ict}. Cisco predicted that by 2023 there will be 14.7 billion devices connect to IoT \cite{cisco2019} with various applications that are generating a huge volume of data \cite{shahraki2018social}. 
Moreover, the popularity of multimedia services over wireless networks is exploding \cite{cisco2019}. Therefore, supporting such a volume of data demands new generations of cellular networks due to the limitation of the previous generations.} 

\textcolor{black}{Since 2019, the fifth-generation (5G) of cellular networks is commercially available in several countries \cite{cave2018disruptive}. 5G uses revolutionary technologies, e.g., higher frequencies, \gls{nfv}, \gls{sdn}, and network slicing and evolutionary technologies, e.g., massive \gls{mimo} to make a significant improvement in data rates, energy efficiency, reliability, and connection density. Meanwhile, 5G network is used in a wide range of applications such as \gls{iot}, smart city, Industry
4.0 \cite{filin20195g}, e-health, wearables, smart utilities \cite{zhang20196g}, etc. Generally, the main 5G service classes include: 1) \gls{embb}, 2)  \gls{urllc}, and 3)  \gls{mmtc}. \gls{embb} is capable of providing high data rates of 1G bits/s for mobile users \cite{zhang20196gg} while \gls{urllc} focuses on the reliability (99.999\%) and latency (milliseconds) of the communication, especially for the applications such as \gls{iiot} and \gls{v2x} \cite{zhang20196gg}. \gls{mmtc} emphasizes on the number of connected devices in \gls{iot} employment (up to 1 million connections per km$^2$) \cite{chai2018energy}.} %Most of these technologies try to automate the procedures of establishing and maintaining the high-performance 5G networks.

However, By taking the present-day and emerging advancements of wireless communications into account, 5G may not meet the future demands for the following reasons:
\begin{itemize}
    \item \textcolor{black}{Given the ever-increasing growth of the deployment of \gls{iot} devices, there is a particular need to improve further the connection density (10 million connections per km$^2$) and coverage of 5G-enabled \gls{iot} networks \cite{jiang2017overview,rm2020load}.}
    \item \textcolor{black}{New emerging services of \gls{iot} such as \gls{xr}, telemedicine systems, \gls{mmi}, and flying cars will challenge the original 5G service classes. To effectively provide \gls{iot} services such as \gls{xr} and telemedicine systems for mobile devices, future mobile networks must simultaneously provide high transmission rates, high reliability, and low latency, which significantly exceeds the original goals of the 5G networks \cite{gupta2020bats,khan20206g,lv20206g}.}

    \item \textcolor{black}{It is expected that future mobile networks will be an ultra-large-scale, highly dynamic, and incredibly complex system, e.g. the massive and heterogeneous devices in the \gls{iot}. However, the architecture of the current wireless networks (e.g., 4G and 5G) are often fixed, and the optimization tasks are defined to cope with specific and identified challenges and services. Hence, the prevailing manual-based optimization and configuration tasks are no longer appropriate for future networks \cite{nayak20206g,zhang20196gg,zhang2020envisioning,tang2019future}.}
\end{itemize}
To deal with the challenges mentioned above, 6G networks are expected to provide new service classes, use new spectrum for wireless communications, enormous network capacity, ultra-low latency communications, and adopt novel energy-efficient transmission methods \cite{yang20196g}.

We aim to present a comprehensive survey on 6G cellular networks by considering a wide ranges of 6G aspects.
Our contributions can be summarized, but not limited, as follows.\textcolor{black}{
\begin{itemize}
    \item Discussing the need of a new generation of cellular networks beyond 5G. 
    \item Providing a detailed review on the existing works on 6G. 
    \item Introducing the 6G \glspl{kpi} and new use-cases e.g., holographic communication and Industrial automation.
    \item Studying various potential enabling technologies that will have important contributions in 6G.
     \item highlighting the potential 6G requirements, challenges, and trends e.g., Green 6G and 3D coverage.
    \item Outlining several 6G future research directions.
 \end{itemize}
}
The rest of this paper is organized as follows. Section \ref{Related Works} reviews the related survey and magazine articles on 6G. In Section \ref{requirements}, we present the requirements and trends of 6G. The research activities and motivation are discussed in Section \ref{section:research activities}. Moreover, in this section, we provide a comprehensive list of 6G \glspl{kpi} and use-cases, as well as new service classes. In Sections \ref{sec:evolutionary} and \ref{sec:revolutionary}, we introduce the important evolutionary and revolutionary technologies respectively that contribute to the future 6G networks. We discuss several 6G challenges in Section \ref{challenges}. Finally, Section \ref{conclusion} draws the conclusion. %The organization of this paper is shown in Fig. \ref{structure}.

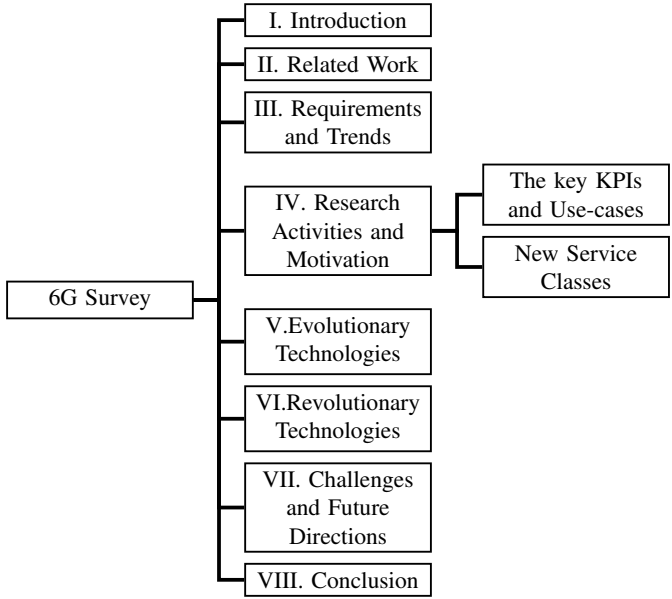
\begin{figure}[t]
\tiny
    \centering
    \caption{The organizational structure of the survey.}
    \label{fig:diagramsolution}
\begin{tikzpicture}[scale=1.78, grow'=right,level distance=0.7in,sibling distance=.03in]
\tikzset{edge from parent/.style= 
            {thick, draw, edge from parent fork right},
         every tree node/.style=
            {draw,minimum width=0.5in,text width=0.5in,align=center}}
           
\Tree 
    [. {6G Survey} 
        [.{\ref{Introduction}. Introduction}
        ]
        [.{\ref{Related Works}. Related Work}
        ]
        [.{\ref{requirements}. Requirements and Trends}
        ]
        [.{\ref{section:research activities}. Research Activities and Motivation }
       [.{The key \glspl{kpi} and Use-cases} ]
       [.{New Service Classes} ]
        ]
       [.\ref{sec:evolutionary}.{Evolutionary Technologies} ]
       [.\ref{sec:revolutionary}.{Revolutionary Technologies} ]
        [.{\ref{challenges}. Challenges and Future Directions}
        ]
        [.{\ref{conclusion}. Conclusion}
        ]
       ]
\end{tikzpicture}
  \label{structure}
\end{figure}
\printglossary[title=List of abbreviations,nonumberlist]

%% file: Sections/related_Work.tex
\section{Related Survey Articles} \label{Related Works}
%The 5G cellular networks and related technologies (e.g., network slicing and \\gls{mmwave}) have been well researched jointly and independently in the existing works; but when it comes to 6G, it only recently has stood at the center of attention. This is mainly due to the fact that the 5G network design is not sufficient to fully exploit this network's potentials and meet the current and future demands. 

Several papers provide a vision of the 6G mobile networks including enabling technologies, potential applications, requirements, and potential challenges. \textcolor{black}{In the sequel, we present a brief overview of the existing survey papers that discussing different aspects of the 6G networks, and compare them with our work.}

The authors in \cite{david20186g} provided a vision of the 6G network and its requirements (e.g., battery lifetime) from a user viewpoint. Meanwhile, the authors envisioned some key enabling technologies for 6G, such as  energy harvesting techniques, the \\gls{ai} and \\gls{ml}. \textcolor{black}{In \cite{9390169} the authors investigated driver factors for 6G (i.e., holographic, massive connectivity, and time sensitive/time engineered applications), the 6G network design principles, and propagation characteristics of the 6G networks.\textcolor{black}{ The authors in \cite{alghamdi2020intelligent} introduced the state-of-the-art of \\gls{lis}, and then 6G.}}

The work in \cite{huang2019survey} discussed the evolution of mobile networks to the 6G network. \textcolor{black}{More specifically, the authors focused on expected architectural changes, e.g., \\gls{3d} architecture and pervasive AI in the 6G networks. They highlight that these changes are essential for cellular networks to satisfy future use cases' demand.} In \cite{saad2019vision}, the authors paint a comprehensive picture of 6G, foreseen applications, emerging trends, enabling technologies, and open research challenges. The authors in \cite{zhang20196g} presented a vision of the 6G for wireless communications and discuss the applications and critical abilities of 6G. They also refer to \\gls{ai} as a key enabling technologies in the 6G era to achieve autonomous network and producing innovative services/applications. As AI-based mobile applications become increasingly popular, the authors in \cite{letaief2019roadmap} discussed 6G networks from an \\gls{ai} perspective. In this paper, the authors pointed out that the 6G networks will be expected to adopt ubiquitous \\gls{ai} solutions from the network core to the edge devices. In \cite{strinati20196g}, the authors highlight the need for a new communication wireless network from technological and societal perspectives. Then, they list several new use case scenarios that are not possible to be efficiently supported by the current 5G networks, such as holographic communications, smart environments, and high-precision industrial manufacturing. A similar work towards envisioning 6G wireless communications has been performed in \cite{yang20196g}. They described the potential 6G requirements and provided a sketch of the latest technological achievements evolving to 6G networks. Furthermore, the authors presented several technical challenges related to 6G and potential solutions.
\begin{table*}[!h]
\centering
\caption{Comparison of Related Survey Articles.}
\label{surveys}
\begin{tabular}{|M{0.84cm}|M{0.6cm}|M{1.4cm}|M{0.6cm}|M{0.7cm}|M{0.8cm}|M{1.2cm}|M{1.3cm}|M{1.1cm}|M{5.5cm}|}\hline
Research & Year & Requirements & Trends & Vision & Service Classes & Architecture & Applications& Challenges& Enabling Technologies \\\hline
\cite{david20186g} & 2018 & \faPlusCircle &\faClose &  &\faCheckCircle&  \faClose &\faClose &\faPlusCircle   &{Optical communication, radio charging}  \\\hline
\cite{huang2019survey} & 2019 & \faPlusCircle & \faClose & \faCheckCircle & \faClose & \faCheckCircle & \faClose & \faPlusCircle& {Terahertz, \gls{vlc}, energy harvesting, molecular communication, blockchain \gls{qc}-assisted, intelligent reflecting surface (IRS)} \\ \hline

\cite{saad2019vision} & 2019 & \faCheckCircle  & \faCheckCircle  & \faPlusCircle & \faPlusCircle  & \faPlusCircle & \faCheckCircle & \faCheckCircle & {Terahertz, \gls{vlc}, edge AI, non-terrestrial technologies, (IRS), energy harvesting }  \\\hline

\cite{zhang20196g} & 2019 & \faCheckCircle & \faClose & \faCheckCircle & \faCheckCircle & \faCheckCircle & \faPlusCircle & \faClose& {Blockchain, terahertz, LISs, molecular communication, QC-assisted, VLC and laser, OAM multiplexing, holographic beamforming } \\\hline

\cite{letaief2019roadmap} & 2019 & \faCheckCircle & \faPlusCircle & \faCheckCircle & \faPlusCircle & \faCheckCircle & \faClose & \faPlusCircle& {AI, big data, AI-powered closed-loop optimization, intelligent communication }\\\hline

\cite{strinati20196g}& 2019 & \faPlusCircle & \faClose & \faPlusCircle & \faClose & \faPlusCircle  & \faCheckCircle& \faClose& {Edge AI, self-optimization, 3D coverage, terahertz and VLC, distributed security } \\\hline

\cite{yang20196g} & 2019 & \faPlusCircle & \faClose & \faCheckCircle & \faClose & \faClose & \faPlusCircle & \faCheckCircle & {Terahertz and VLC, mmWave band, non-terrestrial technologies, multi-mode ultra-massive MIMO, OAM multiplexing, multi-domain index
modulation, AI and big data, heterogeneous networks } \\\hline

\cite{tariq2020speculative}& 2020 & \faCheckCircle & \faClose & \faCheckCircle & \faClose & \faClose & \faCheckCircle& \faCheckCircle& {Pervasive AI, radar-enabled communications, cell-free network, metamaterials, VLC and WPT, energy harvesting, OAM multiplexing} \\\hline

\cite{viswanathan2020communications} & 2020 & \faCheckCircle & \faCheckCircle & \faPlusCircle & \faCheckCircle & \faCheckCircle & \faPlusCircle & \faClose & {AI-empowered air interface,
AI-empowered optimization, sub-networks, hyper-specialized slicing, new cognitive spectrum} \\\hline

\cite{LU2020100158} & 2020 & \faPlusCircle & \faClose & \faCheckCircle &  \faClose & \faPlusCircle & \faCheckCircle& \faCheckCircle& {Terahertz and VLC, synthetic materials, very large scale antenna, blockchain, AI} \\\hline

\cite{9023459} &2020 & \faCheckCircle & \faPlusCircle & \faClose &  \faCheckCircle & \faCheckCircle & \faPlusCircle & \faCheckCircle& {AI/DL and distributed processing, space-air-ground-sea integrated (SAGSI), non-terrestrial technologies, super massive MIMO, advanced multiple access techniques, rate splitting multiple access (RSMA), WPT and energy harvesting} \\\hline

\cite{dang2020should}& 2020 & \faPlusCircle & \faClose & \faCheckCircle &\faCheckCircle & \faClose & \faCheckCircle& \faPlusCircle& {AI, terahertz, operational/environmental intelligence} \\\hline

\cite{9390169}& \textcolor{black}{2021} & \faCheckCircle  & \faClose & \faPlusCircle & \faCheckCircle  & \faClose & \faCheckCircle& \faCheckCircle& {\textcolor{black}{New frequency bands, new physical layer techniques, multiple antenna techniques, intelligent surfaces, multiple-access techniques, AI}} \\\hline

\cite{jiang2021road}& \textcolor{black}{2021} & \faCheckCircle  & \faClose & \faPlusCircle &\faCheckCircle & \faClose & \faCheckCircle& \faCheckCircle& {\textcolor{black}{AI, blockchain, digital twin, IE, communication-computing-control convergence}} \\\hline

\cite{alghamdi2020intelligent}& \textcolor{black}{2021} & \faPlusCircle  & \faClose & \faPlusCircle & \faClose & \faClose & \faPlusCircle & \faCheckCircle & {\textcolor{black}{LIS technology}} \\\hline

Our survey & 2021 & \faCheckCircle & \faCheckCircle & \faCheckCircle & \faCheckCircle & \faPlusCircle & \faCheckCircle& \faCheckCircle& {Terahertz, AI, non-terrestrial technologies, optical wireless technology, 3D network architecture, energy harvesting, quantum communications, LISs, edge intelligence} \\\hline
\multicolumn{10}{l}{(\faCheckCircle: The paper investigated the determined factor, \faPlusCircle: The paper partially covered that factor, and \faClose: The papers did not consider that factor)}
\end{tabular}
\end{table*}

One of the latest studies on 6G networks has been conducted in \cite{tariq2020speculative}. The authors provided their vision of 6G networks and discussed 6G requirements. Moreover, they speculated on the 6G era applications and shed light on the main expected challenges. %Establishing a vision of future mobile networks is the main objective of the work has been conducted 
In \cite{viswanathan2020communications}, the authors presented a list of the 6G requirements derived by the potential future use cases (e.g., \\gls{ar} and \\gls{vr} for industry and biosensors). The authors in \cite{LU2020100158} provided a systematic overview of 6G networks. Their overview covers enabling techniques, potential use case scenarios, challenges, as well as prospects and development of 6G. One of the few articles that well investigate the 6G network's core services has been conducted in work \cite{9023459}.  The authors in \cite{dang2020should} focused on the future mobile network, with the aim of giving a vision for 6G that would guide the researcher in the beyond 5G era \cite{li2020cooperative}. The authors refer to this critical fact that 6G will still be a human-centric network. Regarding this fact, tight security and privacy would be essential requirements of 6G networks. \textcolor{black}{ Last but not least, in \cite{jiang2021road}, the authors conducted a survey covering potential use cases, core services, vision, requirements, and enabling technologies of the 6G mobile systems. }

\begin{figure*}[h]
  \centering
    \includegraphics[width=14cm]{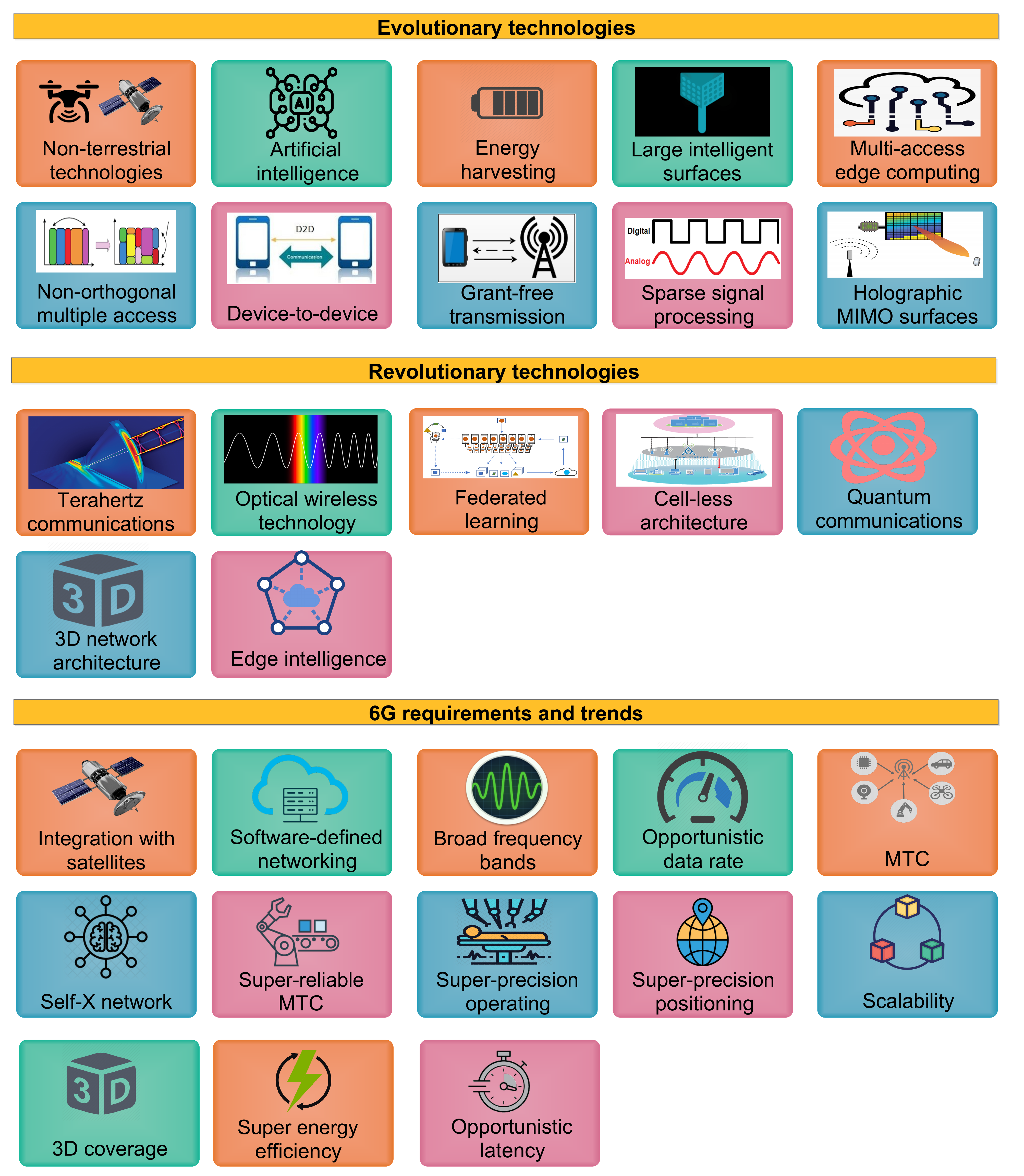}
  \caption{6G vision: Enabling technologies, and requirements and trends}
 \label{figrequirements}
\end{figure*}

\textcolor{black}{To the best of our knowledge, most of the existing survey papers on 6G do not fully cover all aspects of the 6G networks. We have listed the published survey papers in Table \ref{surveys}. The table summarizes the existing survey paper that focuses on
different aspects of the 6G networks such as requirements, trends, vision, service classes, architecture, applications, challenges, and enabling technologies. We represent the topics included in these survey papers and the respective topics in our work. }Compared to the existing literature, the objective of this paper is to give a comprehensive view of 6G networks in various aspects. To this end, we intend to respond to the following fundamental questions:
\begin{enumerate}
  \item Is there any need for a new mobile network generation? If yes, why?
  \item What are the fundamental requirements and trends relevant to the 6G mobile networks?
  \item What are the most recent research activities in the 6G network domain?
  \item What are the critical 6G KPIs and potential applications?
  \item What are the new service classes in the 6G era? And why will these services be needed?
  \item Which are the most critical enabling technologies for 6G and need further study?
  \item Which are the main expected future challenges and open issues in the 6G era that call for significant efforts to resolve?
\end{enumerate}
 We use seven factors for evaluation of the existing works on 6G networks, including requirements, trends, vision, service classes, architecture, applications, and challenges. To the best of our knowledge, the existing papers in the field of 6G networks only partially considered these factors (See Table \ref{surveys}). This work moves beyond the previously mentioned papers and mainly focuses on the fundamental aspects of 6G networks (e.g., requirements, trends, and \glspl{kpi}). Unlike the existing works on 6G that does not fully cover all essential aspects of 6G ( i.e., requirements, trends, visions, new services classes, architectures, applications, challenges, and enabling technologies) (see Figure \ref{figrequirements}), we try to cover all these aspects to an acceptable extent. 

%% file: Sections/requirements.tex
\section{6G Requirements And Trends}\label{requirements}
\begin{figure}[t]
  \centering
    \includegraphics[width=8cm]{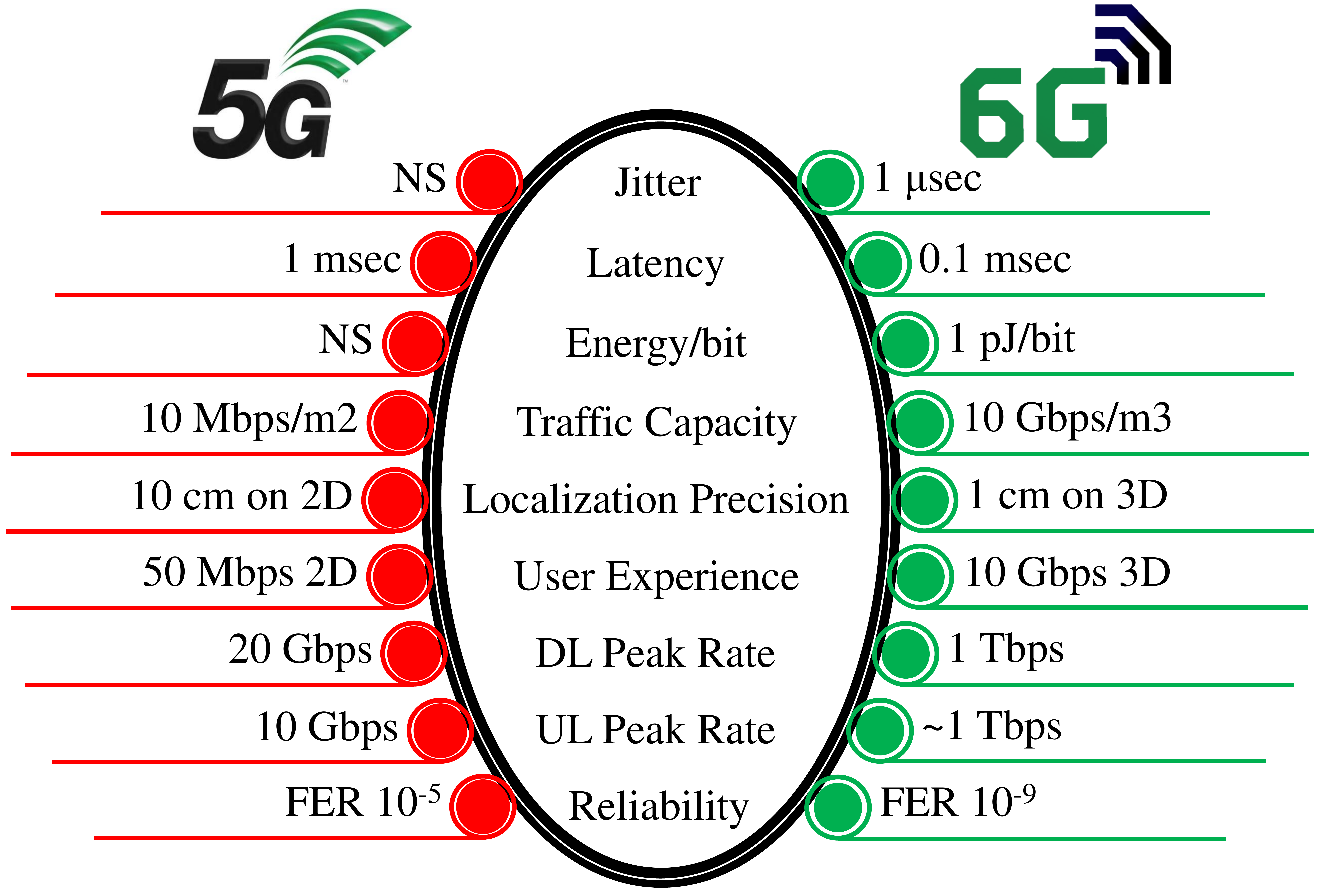}
  \caption{The requirements of 5G and 6G.}
  \label{fig3}
\end{figure}

To tackle the current challenges of the current cellular networks are involved, e.g. \gls{qos}-provisioning in terms data rate, latency and \gls{qoe}, the operators must consider new strategies such as operation in shared spectrum bands, inter-operator spectrum sharing, heterogeneous networks, leasing networking slices on-domain, etc. Ass announced by the corresponding authorities, 6G will need more stringent requirements in compare with 5G as shown in Figure \ref{fig3}. In particular, key 6G requirements and trends are introduced as follows: \cite{piran2019learning}. 
\begin{itemize}
\item \textit{\textbf{Broad frequency bands}}: According to the requirements of the envisioned use-cases of 6G, it is salient that the bands allocated to NR, e.g. sub-6 and mmWave bands, will not be able to support the required \gls{qos} and \gls{qoe}. Therefore, it is predicted that future networks require higher frequency spectrum bands, such as 73 GHz, 140GHz, 1THz, 3THz.  

\item 	\textit{\textbf{Opportunistic data-rate}}: To support the emerging application such as immersive multimedia, a very high and opportunistic peak data rate is required  \cite{sliwa2020towards}. 

\item	\textit{\textbf{Opportunistic latency}}: \textcolor{black}{6G requires the latency to be zero and end-to-end delay to be less than 1 ms. For instance, XR services require a latency to be close to zero in order to improve the QoS \cite{mm}. Furthermore, telepresence require the latency must be lower than sub-millisecond \cite{piran2019learning}.}

\item	\textit{{\textbf{\gls{mmtc}}}}: \textcolor{black}{There will be a number of connected devices that need to be supported by 6G. The current trend, e.g. human-centric solutions will not be efficient due to the network complexity and the sheer number of the connected devices. Therefore, a new trend, machine-centric, will be necessary to support such huge number of devices. However, there will be some critical challenges in this context including scalability, efficient connectivity, coverage improvement, as well as \gls{qos} and \gls{qoe}-provisioning. Moreover, along with high data rate and low-latency communications, 6G use-cases (e.g., mission critical applications) also demand fast connectivity and high reliability/availability (i.e., \%99.999). Towards this end, the 6G networks must comply with strict requirements such as availability, very short latency, and reliability, known as super MTC (sMTC). sMTC will play a critical role in the real-time control process in cyber-physical systems, vehicular communications, and Industry 4.0 operations.}

\item	\textit{\textbf{Self-X network}}: In comparison with the previous generations, 6G networks must be more flexible and robust. Such kind of flexible and robust networks are out of human ability to control. Therefore, to manage such networks, sophisticated \gls{ml} techniques are of vital importance. \gls{ml} techniques are used to support the network autonomy as well as capture insights and comprehension of surrounding environment in which they operate. Hence, with the help of ML algorithms, 6G network will be self-learning, self-reconfiguration, self-optimization, self-healing, self-organization, self-aggregation, and self-protection. 

\item	\textit{\textbf{Super-precision operating}}: It is clear to see that the traditional statistical multiplexing techniques are not adequate to support the future high-precision services and applications. Such as, tele-surgery and intelligent transportation systems. Such services and applications require very high-level and guaranteed precision, e.g., absolute delivery time. To do so, a bunch of new function components are necessary to utilize such as user-network interfaces, reservation signaling, new forwarding paradigm, intrinsic and self-monitoring operation, administration, and management for network configuration.

\item	\textit{\textbf{Super-precision positioning}}: \gls{gps}, uses signal and travel time to find a position. However, the precision of such services are not adequate according to the error that they struggle with, even in order of meters \cite{taki2020indoor}. The future services require very high precision in positioning even in sub-millimeter accuracy. Tele-surgery and tactile Internet are of those services.

\item	\textit{\textbf{Scalability}}: \gls{iot} is going to connect billion of devices ranging from high-end devices to sensors, actuators, smartphones, tablets, wearable, home appliances, vehicles, and many more to the Internet \cite{wozniak20206g}. A huge amount of data is generated via such connected devices. In order to extract the relevant hidden knowledge, \gls{ml} is required. ML-enabled devices can analyze the data and extract the hidden knowledge and hence solve the issue of raw data transmission, which results in improving the network resources utilization.   

\item	\textit{\textbf{Supper energy efficiency}}: 6G devices are supposed to operate in higher frequency band and therefore they need much more energy compared to 5G devices. As an example, energy harvesting is under investigation to overcome the issue of energy efficiency in 6G. 

\item	\textit{\textbf{Connectivity in 3D coverage}}: 6G users will be able to experience beyond 2D in 6G applications such as 3D holographic display \cite{xu20113d}. To achieve such wonderful services, terrestrial and aerial devices will be employed. 

\item	\textit{\textbf{Integration with satellites}}: Satellite communication technologies will be used in 6G to provide global coverage. Telecommunication satellites, earth imaging satellites, and navigation satellites will be integrated in 6G in order to provide localization services, broadcast and Internet connectivity.

\item	\textit{{{\textbf{\gls{sdn}}}}}: network management in 6G needs dynamic and programmatically efficient network configuration. \gls{nfv} enables the consolidation of network instruments onto the servers located at data centers, distributed network devices, or even at end-user premises. Moreover, network slicing offers a cognitive and dynamic network framework on-demand, which can support several virtual networks on top of shared physical infrastructure. 
\end{itemize}

%% file: Sections/Research_Activities_Motivation.tex
\section{6G Research Activities and Motivation} \label{section:research activities}
6G mobile network is envisioned to simultaneously meet future applications' stringent requirements, such as ultra-high reliability, efficiency, capacity, and low latency. Motivated by these foreseen advantages, several research institutes and countries have started research projects on 6G network developments. For example, the \gls{itu} organized a research group for the network 2030, namely FG NET-2030 \cite{clemm2020network}. The central aim of FG NET-2030 is to recognize the suitable set of system technologies to meet the future applications' requirements, such as ultra-massive connectivity and various types of resource requirements. In Finland, the University of Oulu collaborates with NOKIA on the 6G network, \emph{6Genesis} \cite{katz20186genesis}. The project started at the beginning of 2018 to conceptualize the 6G enabling technologies. In 2017,  the European Union (EU) developed a three-year research plan aimed at identifying the essential system technologies of the 6G network \cite{9296115}. One of the funded projects is TERRANOVA, which aims to study and confirm the usefulness of ultra high bandwidth wireless links working in the THz frequency band. The United States, as one of the primary countries in 6G-related research, decided to use the terahertz frequency band for 6G. Towards this end,  Wireless Center at the New York University (NYU WIRELESS)  is designing THz communication channels with up to 100 Gbp/s data rates \cite{rappaport2019wireless}. In China, it launched research on 6G from March 2018 to meet \gls{iot} applications' proliferation in the future \cite{zhang20196g}. \textcolor{black}{LG Electronics company signed an agreement with the Korea Advanced Institute of Science and Technology to develop 6G wireless communication systems in South Korea \cite{khan20206g}. The project's main focus would be on studying THz communications and wireless solutions to achieve 1 Tb/s data transmission speed. Samsung started the study on the 6G networks in June 2019 and released its 6G vision in June 2020 \cite{shafin2020artificial}. This vision discussed different aspects of 6G, including enabling technologies, new core services, and requirements. 6G wireless summit is another research activity towards enabling the next generation of cellular communications. Lapland, Finland, in 2019, has hosted the first 6G wireless summit \cite{rout20206g}. Industry companies such as Ericsson, Huawei, ZTE, and NOKIA are the summit's principal patrons. The 6G wireless summit's main objective is to identify the critical 6G challenges, potential use cases in the 6G era, possible technical requirements, and candidate enabling technologies.}

%Along with these projects, several academic studies have been performed to provide the vision of 6G and help shed light on 6G networks' potential trends. For example, in \cite{9174846}, the authors investigated non-terrestrial-aided cell-free communications for IoT systems in the 6G era.

%In one of the latest papers on 6G, Giordani et al.\ \cite{giordani2020toward} conducted a careful investigation on 6G enabling technologies and relevant applications. They classifieds these technologies into four main categories, including using new spectrum (e.g., THz frequencies and VLC), PHY layer techniques (e.g., out-of-band channel estimation), innovative network architectures (e.g., 3D and cell-less architecture ), and network intelligence (e.g., AI and ML).

%Among all research activities on 6G, higher data rates, ultra-low latency, high connection density, ultra-high reliability are among the most-referred Key Performance Indicators (KPIs). However, there are also other KPIs such as energy and spectral efficiency and delay jitter. In terms of enabling technologies, one can refer to THz communications, AI, optical wireless technology (e.g., VLC), intelligent surface and materials, nonterrestrial communications as the most breathtaking 6G enabling technologies. Nevertheless, technologies such as cell-less architecture, 3D network architecture, Orbital angular momentum (OAM) multiplexing, sparse signal processing, energy-harvesting, grant-free transmission, and MEC are expected to play critical roles in the 6G network.

In the following section, we discuss the main 6G \\glspl{kpi} and foreseen applications.

%% file: Sections/KPIs_Use_Cases.tex
\subsection{The Key 6G \glspl{kpi} and Use-cases} \label{The key 6G KPIs and use cases}
This section discusses the main \glspl{kpi} and characteristics of 6G applications that are envisioned to be widely implemented in the future. A schematic classification of the envisioned services and use-cases for 6G is represented in Fig. \ref{fig5}.
\begin{itemize}
    \item \textbf{\textit{Holographic communication:}} holographic telepresence will be one of the most critical applications of 6G for both profession and social communication \cite{clemm2020toward}. It will enable users to enrich their traditional audiovisual communication with the sense of touch,  while they are in different geographical locations. Holographic communication is a highly data-intensive application, and 5G is unexpected to deal with many holographic communications with absolute reliability. This is mainly due to the fact that holographic imposes strict requirements such as terabits data rate (up to 4 Tb/s), ultra-low latency (sub-milliseconds), and reliable communications. 
 \item \textbf{\textit{Industrial automation:}} It is foreseen that the Industry 4.0 transformation will complete in the era of 6G, i.e., the digital transformation of traditional manufacturing and industrial processes via cyber-physical and IoT systems. The central goal of Industry 4.0 will be to decrease the demand for direct human intervention in manufacturing practices through automatic control systems and networks and communication systems. Toward this end, 6G has to meet tight \glspl{kpi} and requirements, such as a significant level of reliability (i.e., above $1–10^{-9}$), very low latency (i.e., below 1 ms), and multiple connected links
 \cite{berardinelli2018beyond}. Also, some industrial control operations (e.g., industrial \gls{ar} and \gls{vr}) call for establishing real-time communications with very low delay jitter of $1 \mu s$ and Gbp/s peak data rates.
 
 \textcolor{black}{Moreover,  it is envisioned that the 6G network can provide exceptional help for connected robotics and autonomous systems, such as \gls{uav} delivery services, the swarm of self-governing drones, air-ground integrated vehicular network (AGVN) \cite{sun2020surveillance}, and autonomous cars. 6G will enable autonomous vehicles to engage more actively in everyday life, industry, and transportation. More specifically, 6G will fully actualize the large-scale deployment and services of autonomous cars \cite{chowdhury20196g}.}
\item \textbf{\textit{Smart environments:}} The concepts of smart cities and smart homes refer to the smart environments that can significantly increase the quality of life by optimizing the services and operations, resources management, functions automation, services efficiently, monitoring, etc. Accordingly, the 6G network will combine \gls{ict} and an ultra-massive number of smart-physical devices (i.e., \gls{iot} device) to optimize daily life processes such as home security waste management, transportation systems, traffic monitoring, and utilities-related operations, to name a few. Realizing smart environments is one of the critical goals of the 5G network; however, in the 5G era, these applications can partially be realized due to their stringent requirements. One can refer to the high connection density, ultra-high reliability communications (e.g., for transportation systems), tight security (e.g., for smart homes applications), and massive data rates (e.g., for social \gls{xr} and connected cars).
\item \textbf{\textit{E-Health:}} 6G will also offer enormous advantages to the health system through technological innovations such as holographic communication, \gls{ai}, and \gls{ar}/\gls{vr}. 6G will help the health system by excluding time and space boundaries through telemedicine and optimizing health system functions / workflow \cite{zhang2018towards}. Guaranteeing the eHealth services will require satisfying demanding \gls{qos} requirements such as reliable communication (99.99999\%),  ultra-low latency ($<$ 1 ms), and mobility robustness. 

\item \textbf{\textit{Tactile Internet:}} The 5G-enabled IoT systems mostly focus on perception and connectivity. In the future, 6G will provide a more intelligent human-to-machine type of communication for real-time controlling IoT devices, namely tactile Internet. According to \cite{LU2020100158}, the tactile Internet is a wireless communication system that will enable humans and machines to exchange control, touch, and sense data in a real-time manner. The tactile Internet will enable technology for haptics interface and, consequently, possible visual feedback and remote response behavior. The tactile Internet is expected to play a vital role in different services, such as Industry 4.0, everyday life, \gls{cad}, e-commerce 4.0, and other human-machine interaction-based applications. To enable tactile communications in the future, 6G has to provide ultra-real-time response, ultra-low latency and reliable communications. One typical application for tactile Internet is presented in Figure \ref{figsurgery}.

\begin{figure}[t]
  \centering
    \includegraphics[width=8cm]{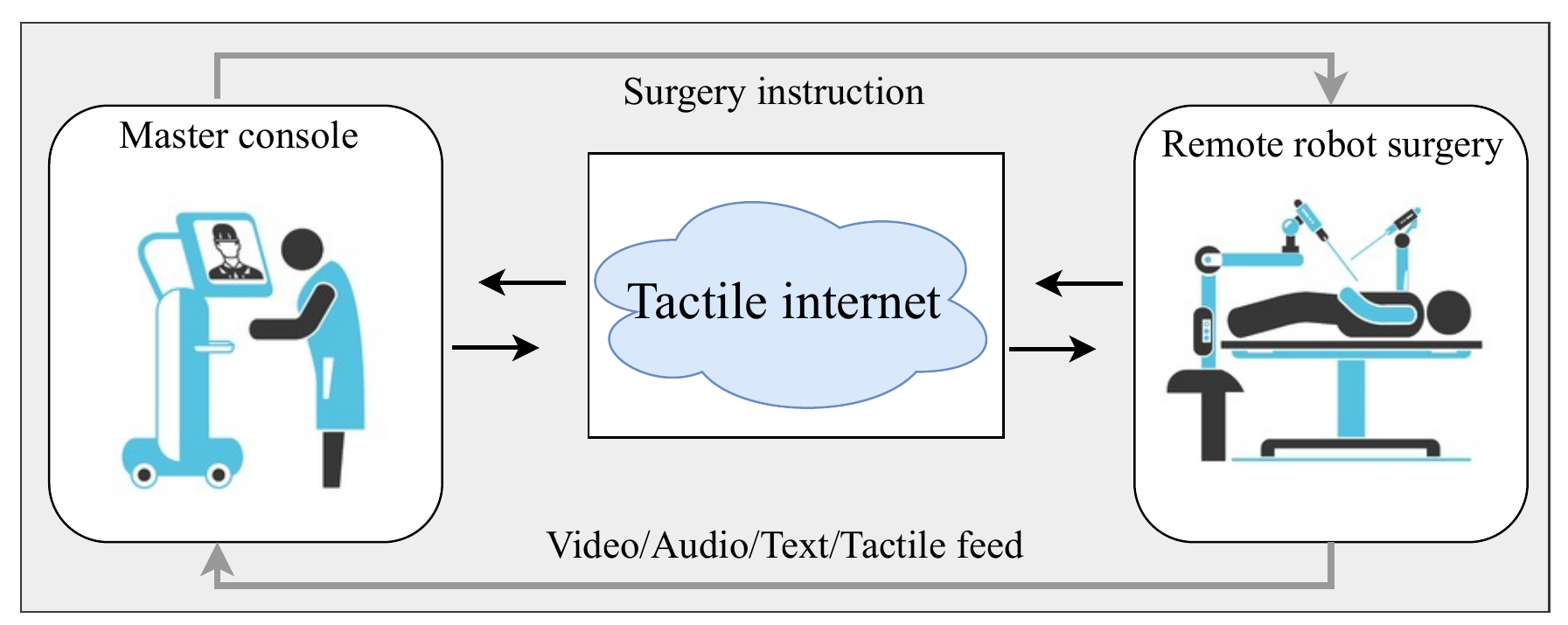}
  \caption{Remote surgery}
  \label{figsurgery}
\end{figure}

\begin{figure}[t]
  \centering
    \includegraphics[width=8cm]{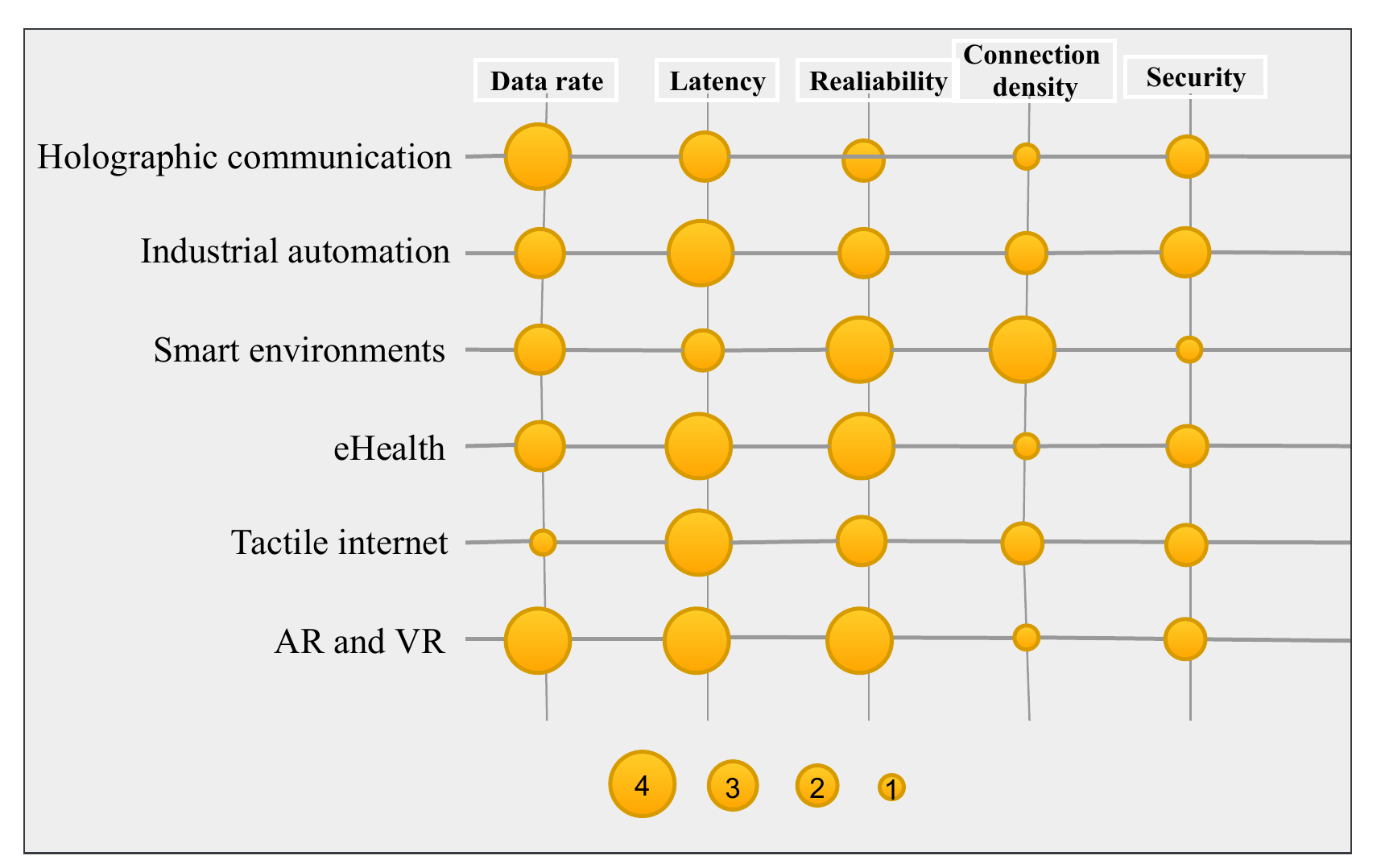}
  \caption{Relative requirement scores for the seven applications – representing graph by ball diameter. Adopted from \cite{FG-NET2030}.}
  \label{fig4}
\end{figure}

\item \textcolor{black}{\textbf{\textit{Private 6G networks:}}}
\textcolor{black}{Different industries and corporations (e.g., factories, airports, oil and gas, health sector, and grids) have employed \gls{iiot} to create an autonomous network for control and monitoring tasks without human intervention. To realize \gls{iiot}, the establishment of the underlying network is crucially important. Traditionally, the underlying network has been created by using both wired (e.g., fiber, industrial Ethernet, and Fieldbus)
 and wireless technologies (e.g., WiFi, WiMAX, and Bluetooth). However, these communication technologies still fail to satisfy the stringent requirements of IIoT applications, such as security, low latency, and reliability. Moreover, the ever-increasing number of IIoT devices, their mobility, and the rise of security and privacy threats motivate the industrial community to replace these technologies with private cellular networks \cite{aijaz2020private}. New launched 5G cellular networks provide an effective alternative to the traditional IIoT underlying network, named the private 5G network, to fulfill stringent communication requirements in industry and other sectors. The private 5G network refers to a 5G new radio technology-based local area network for dedicated wireless coverage in a particular area. The private 5G network can provide distinctive features, such as mobility, positioning, improved security, guaranteed QoS, exclusive capacity, customized service, and intrinsic control that are especially attractive for industrial communication. Another expected evolution path will be the transition from private 5G networks to private 6G networks. This is mainly because many more industries may participate in the competition of adopting custom solutions tailored for their actual use cases, such as industrial automation, warehouse operations, and remote industrial operation.
}

%\item \textbf{\textit{Robotic communications and autonomous systems:}} It is envisioned that the 6G network can provide exceptional help for connected robotics and autonomous systems, such as \gls{uav} delivery services, swarm of self-governing drones, air-ground integrated vehicular network
%(AGVN) \cite{sun2020surveillance}, and autonomous cars. 6G will enable autonomous cars to engage more actively in everyday life, industry, and transportation. More specifically, 6G will fully actualize the large-scale deployment and services of autonomous cars \cite{chowdhury20196g}. Autonomous cars utilize various devices, technologies, and sensors to sense and perceive their environments, such as \gls{lidar}, \gls{gps}, radar systems, and \gls{imu}. The 6G network will establish cellular \gls{v2x} and vehicle-to-server connectivity in a reliable fashion. Regarding the aerial vehicle's connectivity, 6G will provide communication between \glspl{uav} and \gls{gcs}. \glspl{uav} can be used in a broad range of tasks such as smart farming, military usages, business and industry, remote surveillance and monitoring system, to name a few. Furthermore, \gls{uav}-assisted communication is considered as an enabling technology for the future 6G network as we discussed in the sequel \cite{deebak2020drone}.
\item \textbf{\textit{\gls{ar} and \gls{vr}:}} \gls{ar} and \gls{vr} have considered being one of the most distinguished services of the 5G networks. 5G has been adopted the new frequency spectrum (i.e.,\gls{mmwave}) to increase the network capacity, and consequently support data-hungry applications such as AR and VR. AR/VR technologies dramatically affect many research areas and provide new use cases, such as remote surgery, \gls{mmi},  haptic technology, and game technology \cite{van2017challenges}. These use cases will need a different level of latency and reliability. Despite this fact that some of the 5G service classes (e.g., \gls{urllc}) provide high reliability and low latency communications,  some extremely-sensitive use cases (e.g., remote surgery ) will need latency to be shorter than one millisecond, which is not still feasible in the forthcoming 5G network. Moreover, other up-coming VR-based applications such as haptic technology and \gls{vmr} call for massive transmission amounts of real-time data expected to exceed the capability of 5G. Hence, they will need the 6G system to satisfy the end-to-end latency requirements. Figure \ref{fig4}  represents the relative significance of each requirement for the applications.
\end{itemize}

\begin{figure}[t]
  \centering
    \includegraphics[width=8.9cm, height=7cm]{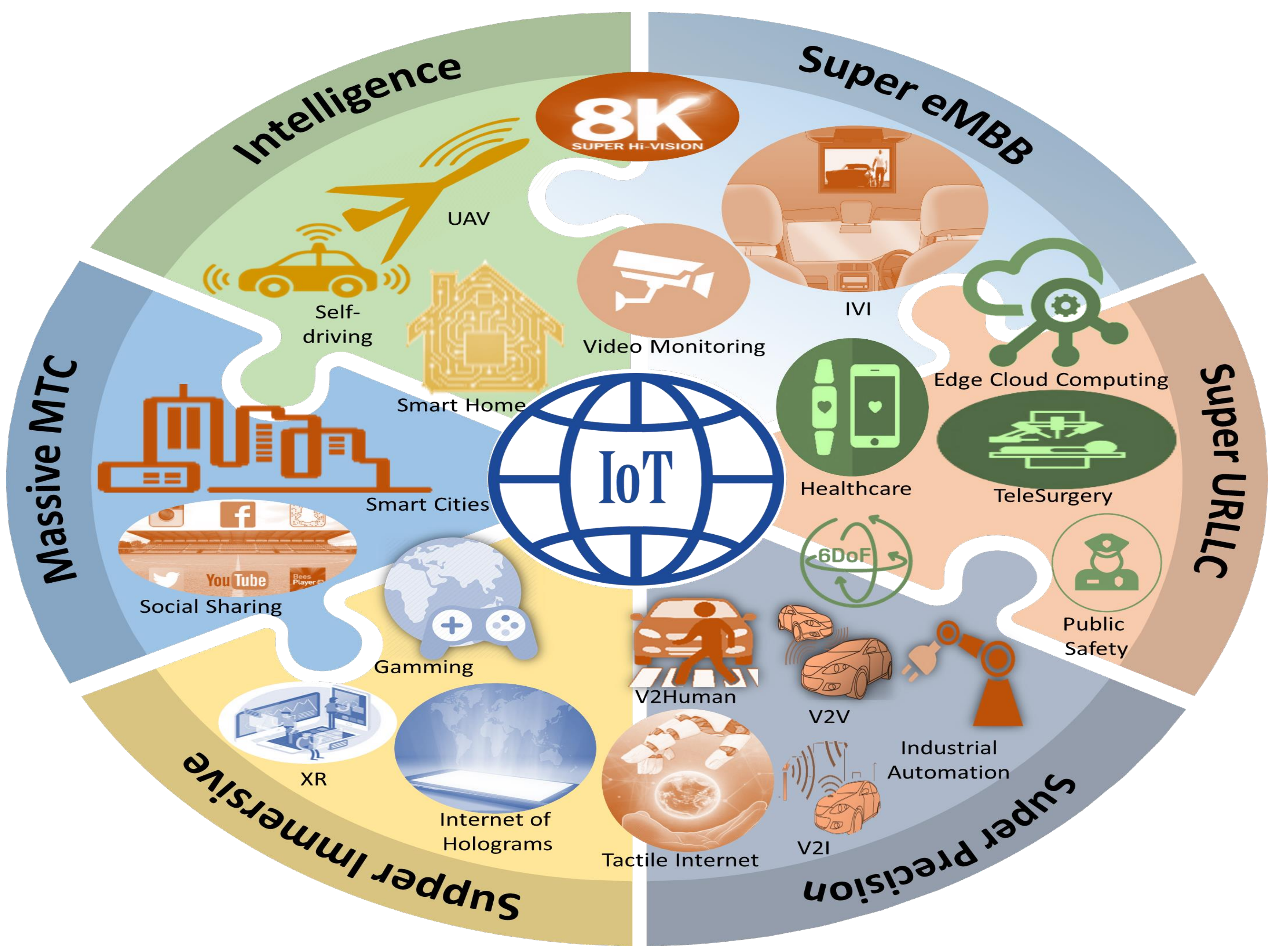}
  \caption{The envisioned services and use-cases for 6G \cite{piran2019learning}.}
 \label{fig5}
\end{figure}

\iffalse
Regarding the mentioned uses cases, a comparison of the main 5G and 6G KPIs is provided in Table \ref{tab:Comparison of 6G with 5G network}.

\begin{table}[]
\caption{Comparison of 6G with 5G network}
\label{tab:Comparison of 6G with 5G network}
\begin{tabular}{|l|l|l|ll}
\cline{1-3}
\textbf{Characteristics} & \textbf{5G} & \textbf{6G} &  &  \\ \cline{1-3}
Rate requirements & 10 Gbps & 100 Gbps- 1 Tbps &  &  \\ \cline{1-3}
\begin{tabular}[c]{@{}l@{}}End-to-End \\ delay\end{tabular} & 10 ms & \textless 1 ms &  &  \\ \cline{1-3}
\begin{tabular}[c]{@{}l@{}}End-to-End \\ reliability\end{tabular} & 99.999\% & 99.99999\% &  &  \\ \cline{1-3}
\begin{tabular}[c]{@{}l@{}}Operating \\ frequency\end{tabular} & \begin{tabular}[c]{@{}l@{}}• Sub-6 GHz\\ • mmWave\end{tabular} & \begin{tabular}[c]{@{}l@{}}THz frequency \\ bands (up to 1 THz)\end{tabular} &  &  \\ \cline{1-3}
\begin{tabular}[c]{@{}l@{}}Connection\\ density\end{tabular} & \begin{tabular}[c]{@{}l@{}}\textgreater 1 million connections \\ per $km^2$\end{tabular} & \begin{tabular}[c]{@{}l@{}}10 million connections\\  per $km^2$\end{tabular} &  &  \\ \cline{1-3}
\begin{tabular}[c]{@{}l@{}}Holographic \\ communication\end{tabular} & partially & fully &  &  \\ \cline{1-3}
\begin{tabular}[c]{@{}l@{}}Extremely-sensitive\\  use cases\end{tabular} & not possible & possible &  &  \\ \cline{1-3}
Autonomous \gls{v2x} & partially & fully &  &  \\ \cline{1-3}
AR/VR & partially & massive scale &  &  \\ \cline{1-3}
\end{tabular}
\end{table}
\fi

%% file: Sections/New_Service_Classes.tex
\subsection{New Service Classes for 6G } \label{New service classes for 6G}
The applications mentioned above and related requirements will lead to new 6G service classes. These service classes expected to refine 5G core service classes, i.e., \gls{urllc}, \gls{embb}, and \gls{mmtc}. In the following, we explain the identified new service classes for the 6G network.

\begin{itemize}
    \item \textbf{\textit{Massive \gls{urllc}:}}\textcolor{black}{5G URLLC refers to communications with high reliability, low latency, and high availability for mission-critical scenarios, such as \gls{iiot} and remote surgery. \gls{urllc} for many devices will be an essential scenario for future communication systems and networks \cite{dhillon2017wide}. Towards this end, 6G has to increase the size of 5G URLLC service to a massive scale, leading to a new service class, called massive URLLC, which combines 5G URLLC with classical \gls{mmtc}. \gls{aid} is one of the foreseeable applications of massive URLLC, in which several important considerations must be taken into account simultaneously, such as motion planning, automated driving, automatic vehicle monitoring, obstacle detection, emergency rescue operations, and so on. Massive URLLC 6G must provide low latency, high reliability, high data rate, massive connectivity, and full mobility at the same time to meet such applications' requirements. 
    \textcolor{black}{To realized massive URLLC, multiple access techniques such as OMA, NOMA, and contention-based multiple access could be promising solutions. By applying OMA techniques,  massive URLLC 6G could experience a linear increase in the required bandwidth along with the rise in the number of devices. Moreover, other multiple access techniques (e.g., NOMA and contention-based) can be used to achieve proper trade-offs among latency, reliability, and scalability \cite{singh2017contention}. Massive URLLC calls for the transmission of massive short-data packets to guarantee time-sensitive 6G applications with excellent resource efficiency and low latency.}}
    \item \textbf{eMBB:}\textcolor{black}{ Furthermore, some other application such as AR, VR, and holographic meetings are promising examples of 5G and beyond 5G applications. Such applications often need high transmission rates (e.g., high-quality video streams), low latency (e.g., real-time interactive instructions), and high-reliability communications \cite{tariq2020speculative}. Moreover, these requirements must also be satisfied in high mobility conditions such as sea and air travel. As a result, a new service class, the so-called enhanced mobile broadband URLLC (\gls{embb} + \gls{urllc}), has been envisioned to allow 6G to support any scenarios subject to the requirements of the high rate rate-reliability-latency. Energy efficiency will be a severe concern for this service class due to its direct effect on reliability and data rate. \textcolor{black}{Compared with the eMBB and URLLC in the 5G networks, this envisioned service class should be highly competent in optimizing mobile communication systems in terms of the handover process, interference, and big data transmission/processing. Furthermore, the security threats and privacy concerns related to enhanced mobile broadband URLLC communication service shall be considered. It seems that the developing resource sharing technique for the coexistence of uRLLC and eMBB services in the 6G networks is one of the most significant technical challenges towards enabling enhanced mobile broadband URLLC \cite{bairagi2020coexistence}. }}
    \item \textbf{\textit{Massive eMBB:}} in Section \ref{The key 6G KPIs and use cases}, we mentioned that tactile Internet would be one of the 6G use cases (e.g., Industry 4.0), which will pose stringent requirements such as high data rates, ultra-low latency, and reliable communications \cite{zhang20196g}. Meanwhile,  to gain tactile perceptions and convert them into digital information, the connection density would often be very high in Industry 4.0-based scenarios(e.g., 100 connections in $m^3$). Hence, massive eMBB will attract lots of interest in the 6G era for improving the operations and functions in large-scale IIoT by supporting massive low-latency connectivity among workers, sensors, and actuators.
\end{itemize}
Beyond these services, the works have been conducted in \cite{letaief2019roadmap} and \cite{sergiou2020complex} envisioned three other service classes for 6G, including Computation Oriented Communications, Event Defined uRLLC, and Contextually Agile eMBB Communications. In \cite{zong20196g}, Zong et al. refer to the fact that advances in industry and autonomous intelligent driving will lead to the appearance new core service classes in the future network, such as \gls{uhsllc}, \gls{uhdd}, and \gls{umub}. Considering foreseen 6G usage scenarios and \glspl{kpi}, a different classification of new service classes will be supported by 6G is introduced in \cite{zhang20196g}. These services include extremely reliable and low-latency communications, ultra-massive machine-type communications, further enhanced mobile broadband, ultra-low-power transmissions, and long-distance and high-mobility communications. Moreover, Human-Centric Services and Multi-Purpose \gls{3cls} and Energy Services have been introduced as new 6G service classes \cite{saad2019vision}.

In the next section, we discuss the technologies that are expected to be integrated into 6G as enabling technologies.

%% file: Sections/Evolutionary_technologies.tex
\section{Evolutionary Technologies of 6G}
\label{sec:evolutionary}
\textcolor{black}{This section and the next one investigate the technologies arising as enablers of the 6G network use-cases and related \glspl{kpi}, discussed in Sections \ref{New service classes for 6G} and \ref{The key 6G KPIs and use cases}. Some of these technologies have already been considered or discussed in the 5G networks; However, due to the technological limitations or market boundaries, they are not commercially available for the 5G networks. The 6G breakthroughs can happen in different layers (e.g., PHY), network architecture, communication protocols, network intelligence, etc. This section is dedicated to evolutionary technologies of 6G. As mentioned above, evolutionary solutions aim to use existing or previously adopted technologies (e.g., MIMO) to realize the 6G networks, where revolutionary solutions aim to exploit novel technologies (e.g., THz communications) to serve 6G. In other words, the revolutionary technologies will fundamentally change different layers of cellular networks (e.g., PHY layer) compared to the 5G networks. Note that many aspects of the revolutionary technologies are still at the step of scientific investigation.
 }
\subsection{\textbf{Non-Terrestrial Technologies}} \label{Non-Terrestrial technologies}
The existing cellular networks based on legacy terrestrial technologies have risen to the challenges of providing extensive wireless coverage to rural areas, the lack of availability/ reliability, and vulnerability to natural and human-made disasters. To address these challenges, the 6G networks will integrate with non-terrestrial technologies, i.e., \gls{uav}-assisted wireless communications and satellite connectivity, in order to afford full coverage and high capacity connectivity \cite{chen2017caching}.
\begin{itemize}

  \item {{\gls{uav}-assisted wireless communications:}}
\glspl{uav} can fly at low altitudes (\textgreater 100 m) and have recently obtained ever-increasing interest thanks to their simplicity and low cost to implement broadband broad-scale wireless coverage during emergencies or act as relay nodes for terrestrial wireless communications. One of the promising foreseen applications of \gls{uav}-assisted communications will be in 6G-enabled \gls{iot} networks. \glspl{uav} can overcome geographical and environmental limitations on wireless communications such as ships on the ocean, deployed sensors in the remote/ isolated regions, and out of terrestrial network coverage areas. It is almost impossible for traditional IoT communication technologies such as \gls{lorawan} and \gls{nbiot} to be applied to such situations.

Regarding integrating \glspl{uav} into the existing and future cellular networks, two possible scenarios can be imagined \cite{zeng2019accessing}. In the first scenario, UAVs can be incorporated into the cellular network as a new category of \gls{ue} that gets services while it flies in the sky, referred to as \gls{au}. It is envisioned that \glspl{au} will be a win-win solution for \gls{uav} technologies and cellular networks because it is a cost-effective solution. More significantly, \glspl{au} can reuse many installed cellular base station (BS) without the need to construct new necessary infrastructures. \glspl{au} will introduce many new UAV-based use cases in the 6G era, such as urban/ road traffic control, search and rescue missions in remote areas, and environmental photography. Moreover, \glspl{au} can be used as a complementary tool to help the positioning systems based on cellular networks. In the second scenario, UAVs can be work as aerial BS or relay nodes to help legacy terrestrial wireless communications by connectivity from the sky. Non-terrestrial BS/relay nodes could deliver enormous benefits compared to terrestrial BS/relays installed at fixed sites \cite{zeng2016wireless}. For instance, aerial BS/relays can be quickly established if desired. This feature is crucially essential for use case scenarios, such as emergencies and unforeseen disasters, and search and rescue purposes. Moreover, compared to the terrestrial BSs/relays, aerial BSs/relays are more likely to establish radio link with terrestrial \glspl{ue} since they are above the ground. Hence, they can make more reliable connections and provide multi-user scheduling and resource allocation for radio access.

Several significant issues, however, posed by UAVs-based communications, such as the different communication \gls{qos} requirements for UAV command/control signals and payload data, severe interference in air-ground communications (uplink/downlink) due to the \gls{los}-dominant channels, and challenges related to UAVs \gls{swap} constraints.
\item {{Satellite communications}}
To achieve ubiquitous connectivity, 6G will integrate satellite communications with the cellular network. High throughput satellite technology will be widely used for establishing broadband Internet connectivity, especially for the remote and out-of-coverage areas of terrestrial communication networks. It is expected that this Internet access service to be competitive with ground-based services in terms of pricing and bandwidth. In the 6G era, satellite-enabled communication is a anticipate alternative for terrestrial and UAV-assisted communication technologies to provide global connectivity for the smart-physical devices on the ground and hence can be utilized for IoT scenarios. Recently, some efforts have been initialized by the \gls{3gpp} to provide satellite communication standards in order to serve upcoming terrestrial wireless communications \cite{hu2020joint}. Motivated by the enormous advantages of satellite-assisted IoT
communications such as communication reliability,  broad coverage, security/protection, fast rural deployment, and long longevity \cite{routray2019satellite}, some satellite communications companies, e.g., Globalstar and Iridium Communications, are involved in designing dedicated satellites for satellite-based IoT networks. One can refer to several applications for satellite-based IoT networks, including mission-critical services, location-based services, navigation systems, healthcare sector, to name a few.
\end{itemize}

\subsection{\textbf{\gls{ai} and 6G}}
Maybe the most influential and recently proposed enabling technology for the 6G network is \gls{ai} \cite{piran2019learning, chen2019artificial}. \gls{ai} has been used in 5G somewhat or very limited, especially ML algorithms such as \gls{dl} and \gls{rl}. Classical ML algorithms also have been applied to 5G, for instance, \gls{svm} and bayesian networks. One can refer to a wide range of \gls{ml} applications in the 5G era, such as \gls{ntc}/\gls{ntp}, \gls{ids}, and \gls{ntma} to name a few \cite{morocho2019machine}. Nonetheless, 6G will fully operationalize AI for different purposes (e.g., intelligent reasoning, decision making, and design and optimization of architecture, operations, and protocols), beyond the classification/prediction tasks. The 6G network is expected to provide ubiquitous AI services from the core to its end devices.

AI can leverage massive training data to learn, provide forecasts, and make decisions, making it a powerful tool for enhancing wireless networks' performance, even when comprehensive system information is not available. Hence, AI can be considered a foundation stone for different aspects, e.g., design and optimization, of future wireless networks, especially the 6G network. It is predicted that the significant impacts of AI on 6G will be in air interface design and optimization. AI methods have been first broadly adopted to meet challenges in the upper layers of communication systems and networks, and achieve remarkable successes \cite{morocho2019machine}. Motivated by these successes, and stimulated by the foreseen KPIs (see Section \ref{The key 6G KPIs and use cases}) and challenges of the next-generation mobile network (see Section \ref {challenges}), AI-based approaches have now been studied also at the other layers of communication systems and networks. In the following, we discuss AI-enabled methodologies and technologies for the 6G network.

\gls{sdn} and \gls{nfv}, also known as network softwarization, are crucial 5G enabling technologies that enable architectural innovation, such as network slicing. Network slicing is one of the key innovative design aspects in the 5G network that allows enormous virtualization capability and, consequently, encourages diverse enterprise business models with a considerable degree of flexibility, more profits for the service provider, and lessen operational costs. However, 6G will be a more complex and heterogeneous network, and hence, network softwarization will not be sufficient. As mentioned, 6G will benefit from new radio interfaces access types such as the THz frequency band and intelligent surfaces. Moreover, the 6G network will need to serve more difficult IoT related functionalities, e.g., sensing, data gathering, data analytics, and storage. When put all together, 6G will need an adaptive, flexible, and intelligent architecture. This calls for further improvement in the current technologies (e.g., \gls{sdn}, \gls{nfv}, and network slicing) to meet the requirements mentioned above. AI-based approaches will present a more versatile network slicing architecture in the 6G era by allowing rapid learning and adaptation to dynamics.

The ever-increasing growth in the volume and variety of data produced in communication networks calls for developing data-driven network operations and planning to adapt to future networks' highly dynamic nature. Leveraging \gls{ml}-based methods for big data analytics is one of the AI applications that can be used for the 6G network. The key applications of AI in 6G networks include \cite{balali2020data}:
\begin{itemize}
    \item Descriptive analytics: Descriptive techniques refer to presenting historical data to easily understand different aspects of the network, such as channel conditions, network performance, traffic profiles, etc.
    \item Diagnostic analytics: Using AI, 6G networks can use the network data to detect the network faults, find out the faulty services, identify the root causes of the network anomalies, and thus enhancing the network performance in terms of reliability and security \cite{kibria2018big,shahraki2020boosting}.
    \item Predictive analytics: Using AI, 6G networks can provide forecasts about the future or unknown events, such as traffic patterns and network congestion, resource availability, and future locations of the users.
    \item Prescriptive analytics: Prescriptive techniques leverage descriptive and predictive analytics to help with the decision about network slicing (e.g., number of slices), virtualization, caching placement problems, and resource allocation.
\end{itemize}

As mentioned earlier, 6G will be a highly dynamic and complex network because of the extremely large-scale, high connection density and heterogeneity. Hence, conventional approaches for wireless network optimization (e.g., mathematical and dynamic programming) will not be applicable for the 6G network \cite{letaief2019roadmap}. As a result of this incapability, it is expected that 6G will benefit from AI-enabled automated and closed-loop optimization. Recent advances in ML methods (e.g., deep \gls{rl}) can build a feedback loop system between the decision-making agent and the cyber-physical systems \cite{abbasi2021deep}. The agent can iteratively improve its performance by receiving the system’s feedback to achieve optimal performance eventually \cite{llll}.

\subsection{\textbf{Energy Harvesting}}
Energy harvesting mechanisms have been incorporated into 5G to meet strict energy limitations affordably and sustainably. These mechanisms can generate electrical power from the external sources for the energy supply of network devices, e.g., BSs and UEs. However, 5G energy harvesting mechanisms currently encounter some challenges, such as the coexistence of these mechanisms with communication protocols and efficiency degradation during converting harvested signals to electricity. Considering the foreseen massive scale of the 6G network, and stimulated by the fact that any sustainable development in communication systems and networks should devote careful attention to energy consumption, 6G will have to develop effective energy harvesting mechanisms and energy-efficient communication techniques. Moreover, it is expected that enabling \gls{iot} in the 6G era through the massive connectivity of low-power and batteryless smart devices. Nonetheless, finding a practical solution (s) to increase batteryless devices' lifetime is a serious issue. Two potential solutions have attracted increasing attention for tackling this issue, including 1) further improve the energy efficiency of low-power devices, and 2) energy harvesting mechanisms and \gls{wipt} \cite{ng2019wireless}. Emerging 6G enabling technologies such as \gls{thz} communications and intelligent surfaces open up ample opportunities to achieve the energy self-sufficiency and self-sustainability vision for the 6G network. For example,  because of its better directionality, the THz frequency band is more efficient than lower frequency bands for \gls{wipt} scenarios. 
\subsection{\textbf{\gls{lis}}}
As we mentioned in Section~\ref{section:research activities}, the spectral efficiency also is one of the 6G \glspl{kpi}. Massive \gls{mimo} is a leading-edge technology to improve the spectral efficiency in which it is possible to serve many UEs in a cellular cell over the identical bandwidth. Besides, the THz frequency band is one of the technologies that promise to help the spectral efficiency. However, in the 6G era, the simultaneous deployment of traditional Massive MIMO technology and THz frequency band can cause several challenges, including high power consumption, extreme complexity in signal-processing algorithms, and increased hardware cost equipment. Using \glspl{lis} for communications will be a solution to alleviate these challenges in the 6G network. Intelligent surfaces are smart electromagnetic materials that can be embedded in our environments, such as buildings, walls, and clothes. These surfaces able to change the reflection of radio waves and expected to lead to the introduction of new communication technologies such as holographic MIMO and holographic \gls{rf} \cite{di2020smart}. The concepts such as smart radio environments, \glspl{ris}, and LISs are intelligent surfaces technology branches, and each of them will bring its advantages. For example, \gls{ris}-assisted systems can improve the UEs' achievable data rate and increase MIMO communication channel rank efficiency.

Among the technologies mentioned earlier, \glspl{lis} gain more attention from academia and industry. \gls{lis} refers to utilizing artificial electromagnetic metasurfaces as large antennas in order to increase the capacity of the network \cite{shlezinger2019dynamic} or to adopt \gls{sal} approach. Indeed, \glspl{lis} can be considered an extension of the traditional massive MIMO technology, but with different array architectures and operating mechanisms. Under using \gls{lis} technology, massive MIMO systems' impressive performance gains can be reached and improve these systems' energy efficiency as LIS' elements do not dependent on any active power source to transmit data \cite{yuan2020potential}.

\subsection{\textbf{\gls{mec}}}
\gls{mec}, formerly recognized as mobile edge computing, is a network paradigm defined by \gls{etsi} and refers to the deployment and execution of distributed computing capabilities, content caching, and network data analytics and network decisions making at the network edge \cite{taleb2017multi}. \gls{mec} will become a primary player in the 6G networks as it can act as an intermediate layer that allows active data analytics, where the data is generated. This paradigm is crucially essential for resource-constrained services/applications \cite{mahmood2019six}. One can refer to V2X communication, improving the energy efficiency and computing offloading of \gls{urllc}, and security and privacy purposes as the prominent use cases related to \gls{mec}. There are several emerging services, such as \gls{ar}/\gls{vr} and \gls{v2x}, that need low \gls{e2e} latency. In this case, \gls{mec} can dramatically decline \gls{e2e} latency by providing edge-based data processing and analytics approaches. Besides, MEC's localized data pre-processing capabilities can reduce the need for sending a considerable amount of redundant or unnecessary data to the cloud data centers. MEC is also expected to be used to efficiently manage network resources (e.g., computational and communication). More specifically, the deployment and operationalization of edge servers, also called MEC servers, at the edge of the network can realize the semi-centralized resource allocation schemes, in which centralized resource allocation techniques can be used to assign the network resources to a cluster of edge devices in the presence of a limited \gls{csi} and with low complexity.

\subsection{\textbf{\gls{noma}}}
Multiple access techniques have always been essential in developing communication systems and networks, and 6G is not an exception \cite{khan2020efficient}. \gls{noma} will become one of the most influential radio access mechanisms for the 5G and 6G cellular networks. \gls{noma} plays an essential role in the implementation and optimization of polar coding and channel polarization methods. As a multiple access technology, \gls{noma} has been proposed for spectral efficiency in 5G/B5G. In comparison to the traditional \gls{oma} technologies, \gls{noma} also represent considerable improvement in terms of security, secrecy capacity, and user fairness \cite{8861078}. NOMA leverages different techniques to provide these improvements, such as \gls{sic} technique and strong/weak users' decoding order. Massive URLLC is envisioned as one of the leading service classes in the 6G networks, where \gls{noma} technologies have remarkable abilities in guaranteeing services such as \gls{mmtc} and \gls{urllc}. Adopting NOMA techniques is crucially vital for current and future mobile networks because they can help improved bandwidth utilization and efficient allocation of resources. Moreover, the MEC convergence with the NOMA, also called NOMA-assisted MEC, can be further studied to enhance the computation service in B5G \cite{li2020joint}.

\subsection{\textbf{\gls{d2d} Communication}}
It is expected that \gls{d2d} communication will be one of the most innovative technologies that help to fulfill the requirements of different emerging use cases in the 6G era \cite{zhang2020envisioning}. Towards this end, \gls{d2d} communication can provide 6G network infrastructure for various D2D-based solutions for \gls{noma}, network slicing, and \gls{mec}, to name a few. Furthermore, it is envisioned that low latency and high-speed D2D communication will be essential for the 6G networks to deal with the limited distance communication because of THz technology in the future \gls{udhn}. %In other words, the main advantage of using \gls{d2d} communication in 6G is addressing the limitations of communication caused by THz technology. In addition to this benefit, \gls{d2d} can be beneficial for \gls{mec}, intelligent network slicing, and NOMA. 

%Although, network densification mechanisms will have to deal with different challenges in the 6G era, such as extreme interference, highly complicated \gls{rrm} techniques, a large amount of signaling, etc, \gls{d2d} communication and \gls{udhn} have brought great benefits for the 5G networks 
%Fortunately, using AI-based solutions is a promising technique to alleviate these challenges and achieve the intelligent automated 6G networks \cite{9023459}. 
%\gls{ai}-empowered \gls{d2d} solutions will provide excellent opportunities and will lead to new challenges in the 6G networks.Basically, using \gls{d2d} communication can enable the network to perform distributed learning in an efficient manner. Various \gls{ml} paradigms \eg federated learning can be achieved based on on-device AI rather than edge nodes. Thanks to growing the storage capacity and computation power of end users devices, they can perform medium-weight \gls{ml} tasks based on local data. \gls{d2d} can connect devices in a close physical proximity to perform distributed machine learning techniques for both ML on network and \gls{ml} for network solutions.

Regarding \gls{mec}, the UEs' spare resources (e.g., computational and communication) can be used to improve network edge computing performance. \gls{d2d} can enable the 6G to use the idle resources by establishing a virtual network infrastructure to manage the resources. Different topology management techniques can be used by \gls{d2d}, e.g., clustering, that can allow the network to use spare resources efficiently \cite{shahraki2020survey}. %As the computational and communication capacity of a single UE is limited and for compute-intensive tasks, the participation of multiple UEs is needed to provide a considerable amount of resources in a low-delay and resource consumption efficient manner. 
In the timeframe of 6G, thanks to THz band D2D technology's employment, the communication between two nearby UEs will be near real-time \cite{abbasi2021synchronization}. Hence, it is expected that the 6G networks will use the capabilities of D2D clusters in edge computing. 

In terms of network slicing, it is envisioned that the D2D-enhanced intelligent network slicing mechanism will encourage telecommunications operators to effectively centralize and combine network resources, such as D2D clusters, private third party, and \gls{plmn} at the edge of the network. Finally, NOMA's utilization as a multiple-access approach for D2D can enable a D2D transmitter to communicate with multiple D2D receivers through \gls{noma}. As a result, the performance gains of D2D will significantly improve \cite{9055054}.

\subsection{\textbf{Grant-free Transmission}} 
Grant-free transmission technology has been introduced as one of the primary trends in future mobile networks \cite{bi2019ten}. Indeed, this technology has been classified as a critical medium access control mechanism for enabling massive IoT connectivity over mobile networks. For the 5G networks, different grant-free transmission techniques have been adopted for \gls{mmtc} and URLLC services; however, these techniques' provided capacity is still restricted \cite{zucchetto2017uncoordinated}. Regarding the ever-increasing growth in the number of smart-physical devices and the popularity of these two services, more efficient grant-free transmission technologies will need to be designed for the 6G networks. Fortunately, the integration of NOMA with the grant-free transmission, GF-NOMA, is a promising solution for the 6G-enabled IoT systems because of NOMA's short delay performance \cite{9097306}. The majority of conventional NOMA techniques considered a centralized scheduling scheme, in which IoT devices are already connected, and different network parameters, e.g., spreading sequences and power control, are pre-determined. However, due to the specific characteristics of mMTC traffic, such as mass uplink communication, small size and periodic data transmission, and various QoS requirements, the conventional NOMA techniques' performance can be highly degraded. In other words, this type of traffic can cause signaling overhead and increase the latency of the centralized scheduler. To deal with this challenge, the grant-free transmission is a viable solution, where the devices can send their data during randomly chosen time-frequency resources in an automatic manner to realize low-latency access and decline signaling overhead associating with the scheduling request. One can refer to signature-based, compressive sensing-based, and compute-and-forward based as the main categories of grant-free NOMA schemes \cite{9097306}.

\subsection{\textbf{Sparse signal processing}}
Sparse sampling, also known as compressive sensing, is a sparse signal processing paradigm that optimally utilizes sparsity in signals for reconstructing them efficiently and with fewer samples. This paradigm's applications have been studied in different aspects of the 5G/B5G networks, including MIMO random access, embedded security, \gls{cran}, channel-source network coding based on the compressive paradigm, etc \cite{wunder2015sparse}. Sparse signal processing algorithms can be used to accurately and effectively recognize active IoT devices in the grant-free transmission approach. One of the main challenges in grant-free transmission, consequently in enabling massive IoT connectivity, is to identify the active IoT devices for data decoding \cite{8454392}. Sparse signal processing is also important for realizing THz communications in the 6G networks. Due to THz channels' sparse nature, compressive sensing methods for sparse channel recovery in THz channels estimation can be used. For example, the work in \cite{sarieddeen2020overview} the applicability of \gls{amp} as a compressive sensing technique has been investigated in THz channel estimation. By leveraging the sparsity feature, the compressive sensing paradigm demonstrates an excellent ability to improve the spectrum and energy efficiency for the future wireless networks and IoT systems. 

\subsection{\textbf{Holographic \gls{mimo} Surfaces}}
One of the key 6G enabling technologies is \gls{hmimos} \cite{9136592}. In the 5G networks, massive \gls{mimo} systems (i.e., BSs with large antenna arrays) have been used to satisfy 5G networks' throughput requirements. Nevertheless, because of various reasons such as energy consumption concerns and the considerable cost of fabrication/operation, it is difficult to fully realize massive MIMO systems. Given the remarkable advances in programmable metamaterials, RISs show enormous potential to deal with massive MIMO challenges and develop the challenging vision for the 6G networks by actualizing seamless connectivity and control of the environment in cellular wireless networks through intelligent software. HMIMOS is expected to improve massive MIMO technology in terms of size, cost, weight, and energy consumption by transforming the wireless network environment into a reconfigurable intelligent entity. Towards this end, HMIMOS may take three different roles, including receiver, transmitter, and reflector. The distinguishing characteristics of HMIMOS (i.e., intelligence and reconfigurability) make it a prospective technology to fulfill the different 6G requirements, including low-latency, low-power, and high-throughput communications. In terms of communications, one can refer to two main groups of applications for HMIMOS, including outdoor and indoor applications. HMIMOS outdoor applications include energy-efficient beamforming, creating connections between the users and BS, PHY layer security,  and \gls{wpt}, where indoor applications are accurate indoor positioning and coverage enhancement in indoor environments \cite{9136592}.

%% file: Sections/Revolutionary_technologies.tex
\section{Revolutionary Technologies of 6G}
\label{sec:revolutionary}
\textcolor{black}{
As mentioned in Section \ref{sec:evolutionary}, 6G enabling technologies can be categorized to evolutionary and revolutionary technologies. In this section, we study the revolutionary technologies of 6G as follows.}
\subsection{\textbf{\gls{thz} Communications}}
Despite the successful 5G deployment with the help from enabling technologies such as the \gls{mmwave} frequency spectrum, the demand for enhancing data rates continues. In this sense, higher frequencies above the terahertz band will be fundamental in the 6G network. The terahertz frequency band is between the mmWave and optical bands, and it ranges from 100 GHz to 10 THz \cite{jornet2011channel}. THz frequency band is promised to provide data rates on hundreds of Gbp/s(e.g., 100 Gbp/s), secure transmissions, extensive connectivity, highly dense network, and enhance spectral efficiency, consequently increase the bandwidth (\textgreater 50 GHz) to meet the requirements of 6G use cases with massive data rates and ultra-low latency. Moreover, the THz frequency band benefits from high-resolution time-domain, which is crucially vital for \gls{sr} sensing technology (e.g., remote sensing) and high precision positioning services (e.g., autonomous driving). Despite these remarkable advantages, multiple unique issues arising from the THz frequency communications. For example, THz links are prone to excessive signal attenuation, rapid channel fluctuation, and severe propagation loss, notably restricting communications over long distances. Besides, to use the terahertz frequency band in commercial communication systems, one should consider engineering-related challenges, e.g., design very large-scale antenna and requiring high  computational power for supporting the extensive bandwidth.

Fortunately, rapid advances in infrastructure and algorithmic aspects of communication systems, such as ultra-massive \gls{mimo} (UM-MIMO), intelligent surfaces, new signal processing methods, and communication protocols, will mature THz communications.

\subsection{\textbf{Optical Wireless Technology}}
Alongside mobile communications based on \gls{rf},  \glspl{owc} will be widely used in the timeframe of 6G. \gls{owc} frequency range comprises \gls{ir}, \gls{vlc}, and \gls{uv} spectrum \cite{chowdhury2018comparative}. \gls{owc} is now being operated since the 4G network. Nevertheless, it will be deployed more broadly to satisfy the requirements of 6G. Among OWC technologies, \gls{vlc} is the most promising frequency spectrum because of the technology advancement and extensive using of light-emitting diodes (LEDs). The OWC in the visible spectrum (380 to 740  nanometers) is generally known as \gls{vlc}, which visible to the human eye.

 For short-range communication distances (up to a few meters), \gls{vlc} technology offers unique advantages over its RF-based counterparts \cite{rehman2019visible}. First, the occupied spectrum by \gls{vlc} systems is free and unlicensed, and they can provide extensive bandwidth (THz-level bandwidth). Second, VLC-based communications do not  \gls{em} radiation and are not seriously interfere with other potential \gls{em} interference sources. This means that VLC communications can be adapted for sensitive EM interference applications such as gas stations, aircrafts, and hospitals. Communication security and privacy is the third advantage of VLC. The transmission medium in a VLC-based network cannot penetrate walls and other opaque obstructions. In other words, the transmission range of the network is limited to indoors. As a result, it can protect users privacy and sensitive information from malicious adversaries. This could be more interesting when we know that about 80\% or more of the time, people tend to stay indoors. Last but not least, VLC can rapidly establish wireless networks and does not need expensive BS since it uses illumination light sources as BSs.
 
 The maximum data rate of OWC is highly dependent on lighting technology. For example, in \cite{lee20154} \cite{tsonev20143} the authors claimed they achieve up to 4Gbp/s data rate with a gallium nitride (GaN)-based LED. Given the technological improvements of LED lamps and related fields, e.g., digital modulation techniques, it is expected that the achievable data rate of VLC will reach hundreds of Gbp/s for the 6G network \cite{viola201715}. It is envisioned that VLC technology will be widely used in different applications, such as intelligent transportation systems (ITS), smart cities and homes, the advertising and entertainment industry, and hospitals.
\subsection{\textbf{\gls{3d} Network Architecture}} \label{3D network architecture}
The currently and previously deployed cellular network architectures are designed for 2-Dimensional (2D) connectivity between network access points and ground-based UEs. Conversely, it is envisioned that the 6G network will integrate the terrestrial and non-terrestrial technologies (see Section \ref{Non-Terrestrial technologies}) to support 3D network coverage. Compared with the fixed 2D infrastructures, the 3D strategy is much more timely and economically efficient (telecommunications operators have to bear the cost of the deployment of dense mobile networks to guarantee massive connectivity), especially when the operators want to quickly provide seamless/reliable/continuous services in rural areas or the case of natural disasters. 3D coverage will also enable communication system for deep-sea and high-altitude. Despite the significant advantages mentioned above, 3D network architecture will pose many challenges that need to be responded before this technology can effectively be used in real-world cellular networks, e.g.,  channel models for air-to-ground communications, trajectory optimization, resource management, topology, etc. \cite{giordani2020toward} \cite{sun2020aviation}. 
\subsection{\textbf{\gls{ei}}}
\gls{ei} or edge AI is another promising computing paradigm that gains enormous interest \cite{chen2020joint} \cite{chen2020wireless}. Big data sources as an enabling technology for learning-based solutions have recently represented a significant shift from the cloud data centers to the ever-increasing edge devices, e.g., smartphones and industrial IoT devices \cite{shahrakicomparative}. It is evident that these edge devices can fuel AI and present several new use cases by providing a massive volume of data. Motivated by the marked shift in big data sources and stimulated by the advantages mentioned earlier, there is an imperative action to push the AI solutions to the edge of the network to exploit the edge big data sources' potential entirely. Nevertheless, providing AI solutions at the network edge is not a trivial task because of the issues related to performance, data/user privacy, and cost. The traditional approach is to transfer the data generated by the edge devices to the cloud data centers for processing and analytics to deal with these issues. Clearly, transferring such a considerable amount of data will bring monetary/communication costs and cause delays. Moreover, data protection and privacy-preserving can also be significant concerns in this scenario. On-device data analytics is proposed as a remedy, in which AI solutions can be run on the edge devices to process generated data locally. Nonetheless, in this alternative, lower performance and energy efficiency are expressed as primary concerns \cite{ogudo2019device}. This is mainly because most AI solutions need immense computational power that significantly exceeds the capability of power- and resource-constrained edge devices. \gls{ei} has been emerged to tackle the issues mentioned above. \gls{ei} is the combination of edge computing and AI, which promises to provide tremendous advantages compared to the conventional approaches based on cloud, such as privacy-preserving, low-latency, efficient energy consumption, cost-effective communications, etc \cite{wang2020convergence}.

\begin{figure}[h]
  \centering
    \includegraphics[width=0.4\textwidth]{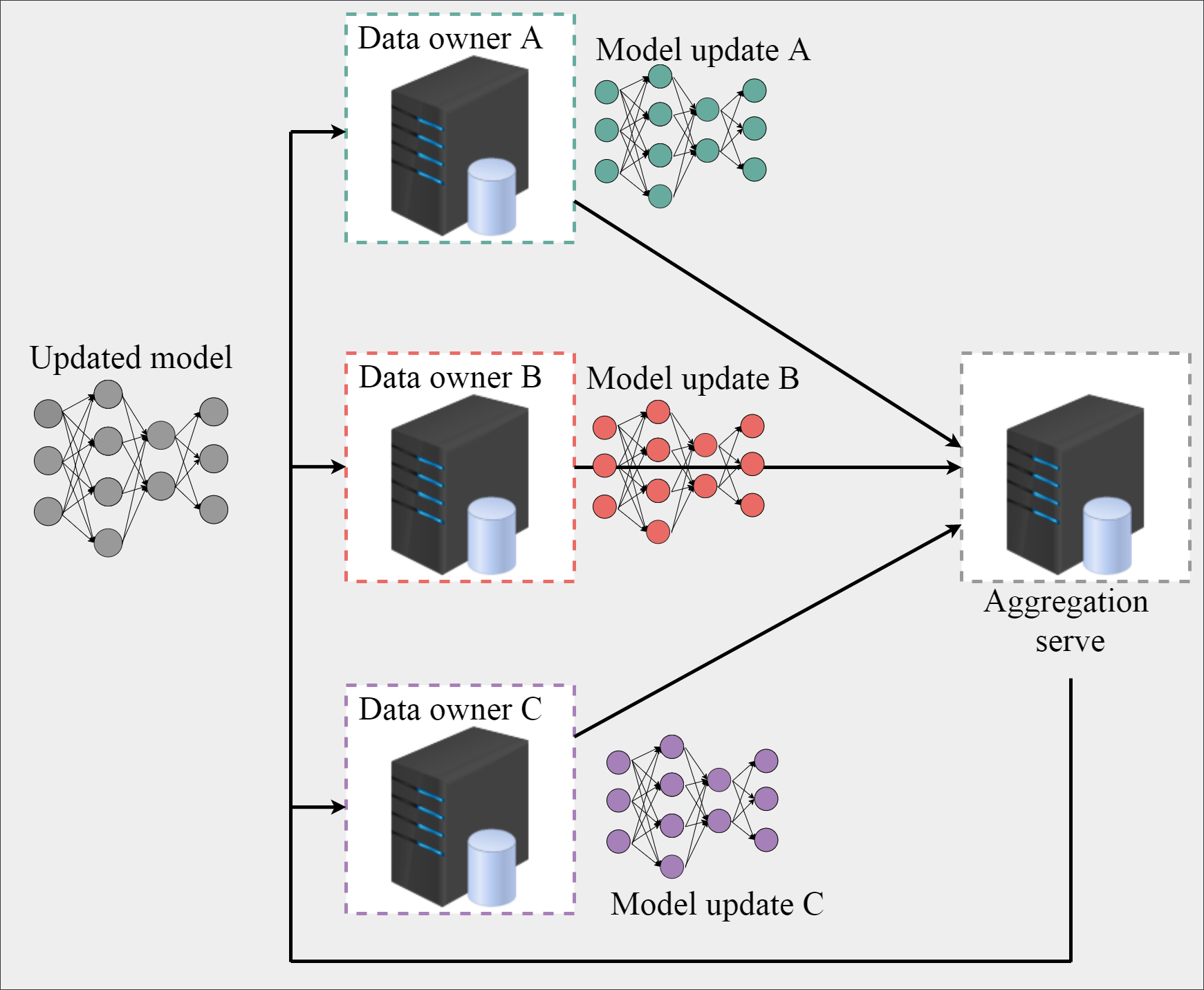}
  \caption{A typical federated learning architecture.}
 \label{fig_federated}
\end{figure}

It is generally believed that the 6G networks will adopt ubiquitous AI solutions from the network core to the edge devices. Nevertheless, conventional centralized ML algorithms need the availability of a large amount of centralized data and training on a central server (e.g., cloud server or centralized machine), which will be a bottleneck in the future ultra-large-scale mobile networks \cite{9170905}. Fortunately, as an emerging distributed ML technique, FL is a promising solution to deal with this challenge and realize ubiquitous AI in the 6G networks. FL is an ML technique in which creating ML models do not rely on storing all data to a central server where model training can occur. Instead, the innovative idea of FL is to train an ML model at each device (participant or data owner) where data is generated or a data source has resided, and then let the participants send their individual models to a server (or aggregation serve) achieve an agreement for a global model (See Figure \ref{fig_federated}).

FL can alleviate privacy and security challenges associated with traditional centralized ML algorithms, as well as guarantee ubiquitous and secure ML for the 6G networks \cite{liu2020federated}. The centralized ML technique is contradictory to ubiquitous ML services, which are promised by 6G. This is mainly due to the data collection- and data processing-related overheads in centralized ML techniques. As a result, distributed ML techniques, mostly FL, in which all training data is located in remote devices locally, are required for future communication systems. FL is showing itself to be an accelerator for the extension of privacy-sensitive applications/services. However, despite the considerable potential advantages of FL for the 6G networks, FL is still in its infancy and encounter various challenges for fully operationalize in the 6G networks.

The main challenges facing FL in the 6G era include significant communication cost for model updating and aggregation, privacy concerns associated with gradient leakage attacks and membership inference attack \cite{wei2020framework}, security concerns resulted from heterogeneous and various data owners (e.g., data poisoning attacks), and the model training and inference concerns caused by the ultra-large-scale of 6G networks. It is envisioned that an enormous number of heterogeneous devices and communication technologies will be deployed in the 6G networks; hence, it is crucially important to improve the communication efficiency in FL algorithms to reduce the number of times the aggregation server gets gradients from the participant devices. Most importantly, it is necessary to develop privacy-enhancing mechanisms in FL as the current techniques proposed for improving privacy in FL, e.g., homomorphic encryption and secure multiparty computation, can not deal with the attacks mentioned above \cite{li2020federated}.

\subsection{\textbf{Quantum Communications}}
Motivated by the great potential of parallelism showed by \gls{qc}, the \gls{qc}-assisted communications field has gained lots of interest. This is mainly due to the fact that quantum communications has a strong potential to meet the stringent requirements of 6G such as  massive data rates, efficient computing, and strong security \cite{gyongyosi2019survey}. Toward this end, technologies such as \glspl{qoc}, quantum optical twin, quantum communication, and \gls{qkd} have been investigated in the literature \cite{manzalini2020quantum} \cite{wilde2012quantum}. The main idea behind \gls{qc}-assisted communications is that \gls{qc} uses photons (or quantum fields) to encode the data in the quantum state (or qubits) and transmits qubits from a quantum emitter to a quantum receiver. Using qubits in communications brings enormous advantages, such as communication security, high-speed and low transmission losses in optical and radio mediums, lessening the chance of decoherence, etc. Moreover, QC-assisted communication shows excellent potentials for long-distance communications. More specifically, quantum repeaters can be used at long distances to divide the communication link into multiple shorter middle segments and then correct errors such as photon loss and operation errors in these segments \cite{ruihong2019research}. 

Several works have practically implemented the applications of quantum-based technologies in communication systems and networks. For example, QKD's capabilities for building a quantum access network have been investigated in \cite{frohlich2013quantum} \cite{cai2019quantum}. Besides, the works have been conducted in \cite{rubino2019experimental}\cite{procopio2015experimental} focused on implementation and testing quantum switches.

AI is another field that will be revolutionized by QC. The currently available AI techniques are quite expensive in terms of energy, time, and resources, especially \gls{dl} models. This is mainly due to the fact that these techniques use traditional Boolean algebra-based transistors for processing a massive amount of data, unusually DL models. In other words, the technological advancement in chipsets does not grow at the same pace as the AI techniques growing. Fortunately, computing systems based on quantum principles are a promising solution to tackle this problem as these systems are significantly faster and more energy-efficient than their traditional ancestors.

\subsection{\textbf{Cell-less architecture}}
Cell-less communication architecture, also known as cell-free, has been proposed to deal with performance degradation poses by the cellular networks' handover process \cite{8246848}. Under this architecture, a UE can communicate with cooperative BSs (or \gls{ap}) through coordinated multipoint transmission and reception techniques instead of connecting to a single BS. Establishing cell-less communications can enhance connectivity and lower the latency induced by the handover process. Cell-less communications will be inevitable in the 6G era due to the fast deployment of heterogeneous communication systems and using several frequency bands, where UEs will transfer from one network to another network without requiring doing handover process \cite{chowdhury20196g}. The UEs will then choose the best link from the available heterogeneous links (e.g., \gls{thz}, \gls{mmwave}, and \gls{vlc}) in an automated manner. As a result, the traditional handover process issues, such as data loss and handover delays/failures, can be alleviated and achieve better QoS. In other words, cell-less communications will ensure UEs' seamless mobility without overhead because of the handover process.

%% file: Sections/Future_Challenges.tex
\section{6G Challenges and Future Research Directions}\label{challenges}
A few potential critical open issues for future research work in the 6G networks are presented in this section. 
\subsection{Energy Consumption Challenges}
In the era of 6G, an unprecedented number of low-power IoT devices and battery-less smartphones will serve over 6G to realize \gls{iot}. As a result, super energy-efficient techniques will be fundamental to guarantee the \gls{qoe}. The energy-related issue is much more notable in smartphones as the current battery life for smartphones without charge is almost one day, which will be problematic for the development of cellular communications. To deal with this challenge and comprehensively improve the 6G network performance so as to serve more end devices, designing effective power supply mechanisms with novel signal processing techniques is necessary. Several energy supply techniques, especially wireless energy harvesting-based techniques, can be adopted in the 6G network. In particular,  the design of low-density channel codes and energy efficient modulation techniques are useful. Energy-efficient communication paradigms also can be investigated, such as symbiotic radio (SR). Furthermore, energy management optimization in the future cellular network is another promising technique to create a trade-off between energy demand and supply dynamically.

\subsection{AI-related Challenges}
It is undoubtable that \gls{ml} is an integral solution in 6G, but there are some challenges that should be considered. AI tasks usually generate a heavy computational load and are often designed, trained, and used at servers with task-customized hardware. Considering the rapid proliferation of smart-physical end devices, it is envisioned that a vast number of AI applications will be designed and used by these devices at the network edge. However, given the current mobile networks, user privacy, data security, wireless link capacity/latency are the main concerns of mobile AI-based applications. Regarding privacy and security, the emergence of decentralized machine learning techniques is a promising possibility to preserve the users' privacy and protect sensitive data. FL tries to create a joint ML model by training an algorithm on the data placed at several distributed sites. In FL, each participant (data owner or client) trains a local model and sends the model weight updates to a central server (a.k.a. aggregation server) instead of sending raw data. FL offers multiple advantages. For example, it preserves the users’ privacy and protects the security of data.  Moreover, FL allows various participants to train an ML model in a collaborative manner to achieve a better model compared to what they can achieve alone.
\textcolor{black}{
\gls{ml} can be integrated by 6G in two aspects including:
\begin{itemize}
    \item \gls{ml} on 6G networks: In this aspect, distributed \gls{ml} techniques \eg FL and split learning are performed on 6G network infrastructures for especial tasks \eg \gls{ei}.
    \item \gls{ml} for 6G: In this aspect, \gls{ml} techniques are used as decision-makers for networking solutions to provide automated network infrastructure establishment and maintenance. In a more depth aspect, the \gls{ml} solutions for 6G can be performed in a distributed manner, although most of existing solution in this aspect working as a centralized model \cite{park2019wireless}. One of the most important aspects of \gls{ml} for 6G is resource management.
\end{itemize}
Nevertheless, both aspects are resource-hungry tasks, but they have two goals. \gls{ml} on 6G networks try to allocate more idle resources to \gls{ml} tasks in an efficient manner, but an optimal \gls{ml} for 6G solutions can establish the 6G network with the lowest resource consumption. To achieve both goals, available resources should be separated between these two aspects in an efficient manner that is a complex task. In other words, it is unaffordable to use a big volume of available resources for managing 6G networks and its available resources as they finally should be allocated to \gls{ml} tasks for especial purposes. }

\subsection{Terahertz Communications Challenges}
THz communications are expected to be a critical enabling technology for the 6G network. However, as we mentioned, THz communications are facing some challenges, notably severe propagation loss and constrained communication over long distances. Hence, research communities need to work jointly together to deal with these challenges and realize THz communications. Fortunately, several research activities are open-ended, such as THz wireless transmission by Fraunhofer HHI, Communications, Sensing at Terahertz by NYU WIRELESS, and ICT-09-2017 Networking research funded by Europe Horizon 2020.

Due to the nature of THz communications' features, the design of THz MAC also poses many challenges, including serious deafness problems, complicated network discovery and coupling operations, and the need for designing efficient concurrent transmission scheduling techniques. These challenges and many others are tackled in work have been conducted by Han et al. \cite{yuan2020potential}. Hardware design is another challenge in THz communications. More specifically, real-world THz communications deployments call for innovations in circuit and antenna design, as well as miniaturizing current big high-frequency transceivers. For instance, the antenna size to support joint communication in mmWave and THz bands may vary (from nanometers to micrometers) and need to be redesigned.
  
\subsection{3D Coverage Challenges}
As mentioned in Section \ref{3D network architecture}, the 6G network will support 3D network coverage because of integrating the terrestrial and non-terrestrial technologies (e.g., UAV-assisted and satellite communications). However, this calls for collaborative research on different aspects of 3D network architecture. First, air-to-ground 3D channel modeling and measurement for communications are required. Second, novel topology optimization and network planning methods (e.g., for the deployment of non-terrestrial BSs/relay nodes) must be developed. Finally, novel network optimization tools and techniques for energy efficiency, mobility management, trajectory, and resource management in 3D network architecture are required.

\subsection{Security Challenges}
Given the new service classes, developed threat landscape, raised privacy concerns, and new wireless communication methods (e.g., non-terrestrial technologies), security/privacy is expected to be a critical issue for 6G. Moreover, network virtualization and softwarization in the 6G era will cause network security/privacy boundaries to gradually fade. As a result, the security defects induced by network architecture have become more and more notable. The increased and closer integration of big data analytics techniques, AI, and EI may also introduce data security risks at the network's edge. The conventional independent security approaches will not be practical for internal network security risks (network- and access-side) posed by enabling technologies, new use case scenarios, and service classes. Hence, it is crucially important to develop the conventional approaches and think about new security methods \cite{9042251}\cite{clemm2020network}.

In the 6G era, new security techniques, e.g., integrated network security, will complement traditional physical layer security techniques, especially if they consider the requirements such as tight security levels and low complexity. Towards this end, the extension of some of the previously proposed PHY layer security methods can be used for 6G. For example, PHY layer security mechanisms for mmWave communications can be adopted for Terahertz communications. As mentioned in \cite{9042251}, one can refer to authentication-, access control-, communication security-, data encryption-related technology as the key security enabling technologies for 6G.

Another envisioned security challenge in 6G is related to the continuing growth of IoT devices proliferation. At present, the most prevalent IoT communication protocols are including 6LoWPAN, short for IPv6 over Low -Power Wireless Personal Area Networks, IEEE 802.15.4, LoRa, and . These communication technologies use cryptographic algorithms such as \gls{ecc} and \gls{aes} as security mechanisms. However, with the emergence of new communication technologies, e.g., QC-assisted communications and the growth in end devices' computational power, the traditional IoT protocols will not be secure \cite{9003618}.

\subsection{Tactile Internet Challenges}
Incorporating control, touch, communication, and sensing capabilities into a shared real-time system pose a crucial issue in accomplishing the tactile Internet. Despite using virtualization and softwarization technologies and \gls{mec} to achieve the low latency requirements, the tactile Internet is still in its infancy and further research is needed to solve some open technical challenges, such as physical layer challenges. Moreover, to alleviate signaling overhead and air interface latency, taking into account several other issues, including optimal waveform selection algorithms, robust modulation techniques, intelligent control plane, Control and User Plane Separation techniques, should is essential. Scalable routing mechanisms and adaptive network coding schemes is also worthy further research as they can reduce end-to-end delay. Besides, security flaws are among the main concerns about tactile Internet-based use case scenarios. Hence, providing an effective security mechanism to deal with malicious activities is crucial for realizing the tactile Internet.

\subsection{Random Access Protocols}
The proliferation of IoT devices calls for providing wireless solutions that can transfer data in a reliable/energy-efficient/spectral-efficient manner. However, this is not a trivial task, mostly when an IoT system consists of a vast number of IoT devices. This is mainly due to the fact that the IoT devices send/receive data packets sporadically, and unpredictably, and consequently makes it challenging to design an effective resource allocation mechanism. \gls{ra} protocols have been proposed to deal with this challenge, where they can decline the cost of communication in terms of wireless resources. The majority of RA protocols that have been employed in the existing wireless solutions, such as 5G and \gls{lorawan}, are not optimal in the networks with a massive number of nodes granting access. To help deal with this situation, the cooperation opportunities between modern \gls{ra} methods with technologies such as \gls{noma}, \gls{ofdm}, massive \gls{mimo}, and sparse signal processing would lead to more efficient wireless systems \cite{de2017random}.

\subsection {\textcolor{black}{Privacy concerns in the future smart services}}
\textcolor{black}{The 6G networks are envisioned to offer ubiquitous and unlimited network access for many users and MTC devices. This seamless connectivity is a prominent supporter of future smart-based services such as smart environments, homes, health, industries, cities, utilities, government, etc. As an example, for a given smart-based service, one might refer to a smart lighting system for a home, which can provide a more efficient way of lighting in terms of energy consumption and cost. However, such systems will use/share sensitive and private information, e.g., occupancy time, household information, habits, and preferences. Indeed, using this de-identified data by a provider to deliver smart services may be a double-edged sword in many scenarios rather than absolutely bring benefit. Given the technological advancements and the availability of rich data flows, another privacy-related challenge is the need for a clear definition of de-identified data. This is especially important when it comes to introducing measures for determining the privacy and sensitivity level of data in a dataset on any occasion. Integrating the blockchain technology can be considered as a potential solution to improve the privacy of 6G, but to the best of our knowledge, there are few efforts to apply the blockchain in 6G \cite{hewa2020role}. 
}

\subsection {\textcolor{black}{Green 6G}}
\textcolor{black}{6G is expected to integrate terrestrial wireless networks with space, aerial and underwater communications to realize connectivity in 3D coverage. Indeed, the primary objective of 6G is to ensure anytime anywhere network connectivity for a vast number of IoT devices and users. These devices/users impose diverse QoS requirements, need for handling massive/heterogeneous amount of traffic, and have very low power consumption through design and building energy-efficient wireless communication protocols, computing, and transmitter/receiver technologies. Besides the trends and innovative technologies (i.e., the evolutionary and revolutionary technologies) that we mentioned throughout the paper, green communication and green computing will be among the next generation of wireless networks' primary goals. This is mainly important for reducing overall power consumption and operating costs, consequently can bring positive effects on environmental and business aspects. Towards this end, making a shift from self-organizing networks to  is an increasing trend in the past few years that can be applied in 6G networks.
 }
\subsection{\textcolor{black}{\gls{nmo}, \gls{ntma}, and 6G}}
\textcolor{black}{6G is considered one of the most important network infrastructures for \gls{iot} networks. Due to the \gls{iot} applications, massive-scale \gls{iot} networks make significant challenges in \gls{nmo} and \gls{ntma} techniques \cite{shahraki2021tonta}. As the former techniques are used to organize the network infrastructure \eg fault management and network configuration management \cite{shahraki2020survey}, the latter methods are broadly used to evaluate networking performance in different aspects \eg performance management and security management. Both techniques can highly affect the \gls{qos} and \gls{qoe} which are the crucial factors in 6G. To the best of our knowledge, there is a shortage of effort to overcome these challenges in 6G, especially in complex 6G network infrastructures \eg \gls{d2d}-enabled 6G and massive-scale \gls{iot} based 6G networks. During the last decade, \gls{ml} techniques are proposed to overcome the challenges of \gls{nmo} \cite{ayoubi2018machine} and \gls{ntma} \cite{abbasi2021deep}, but there is a considerable lack of research in these fields for 6G that can be considered as a future research direction.  }

%% file: Sections/Conclusion.tex
\section{Conclusion}\label{conclusion}
%Research on the 6G networks is still in its infancy and its planning stages. 
\textcolor{black}{In this paper, we studied 6G as the next communication paradigm for \gls{iot}. We have first discussed the need for 6G
networks. Then, we have introduced the potential 6G requirements and trends, as well as the latest research activities related to 6G. Furthermore, the key performance indicators, applications, new services, and the potential key enabling technologies
for 6G networks have presented. Finally, we have presented several potential unresolved challenges for future 6G networks.}